\documentclass[11pt]{article}
\usepackage{comment}
\usepackage[margin=2.5cm]{geometry}
\usepackage{amsmath,amssymb,extarrows,mathtools,graphicx,subfigure,setspace,bbold,physics}
\usepackage{cite}
\usepackage{slashed}
\usepackage[dvipsnames]{xcolor}
\usepackage{hyperref}
\hypersetup{colorlinks=true, linkcolor=blue, citecolor=magenta, linktoc=page}
\usepackage{tikz}
\usepackage{color}
\usepackage{arydshln,leftidx,mathtools}
\usetikzlibrary{decorations.pathreplacing}
\numberwithin{equation}{section}
\newcommand{\be}{\begin{equation}}
\newcommand{\bea}{\begin{eqnarray}}
\newcommand{\eea}{\end{eqnarray}}
\newcommand{\ba}{\begin{align}}
\newcommand{\ea}{\end{align}}
\newcommand{\ee}{\end{equation}}

\usepackage{chngcntr}
\begin{document}

\begin{titlepage}
\thispagestyle{empty}

\begin{flushright}
IPM/P-2021/013\\
\end{flushright}

\vspace{.4cm}
\begin{center}
\noindent{\Large \textbf{Entanglement Wedge Cross Section in Holographic
Excited States}}\\
 
\vspace*{15mm}
\vspace*{1mm}
\vspace*{1mm}
{Mohammad Sahraei${}^{\ast}$, Mohammad Javad Vasli${}^{\ast}$,\\ M. Reza Mohammadi Mozaffar${}^{\ast, \dagger}$ and Komeil Babaei Velni${}^{\ast, \dagger}$}

 \vspace*{1cm}

{\it  ${}^\ast$ Department of Physics, University of Guilan,
P.O. Box 41335-1914, Rasht, Iran\\
${}^\dagger$ School of Physics,
Institute for Research in Fundamental Sciences (IPM),
P.O. Box 19395-5531, Tehran, Iran
}

 \vspace*{0.5cm}
{E-mails: {\tt sahraei@msc.guilan.ac.ir, vasli@phd.guilan.ac.ir, mmohammadi@guilan.ac.ir, babaeivelni@guilan.ac.ir}}%

\vspace*{1cm}
\end{center}

\begin{abstract}
We evaluate the entanglement wedge cross section (EWCS) in asymptotically AdS geometries which are dual to boundary excited states. We carry out a perturbative analysis for calculating EWCS between the vacuum and other states for a symmetric configuration consisting of two disjoint strips and obtain analytical results in the specific regimes of the parameter space. In particular, when the states described by purely gravitational excitations in the bulk we find that the leading correction to EWCS is negative and hence the correlation between the boundary  subregions decreases. We also study other types of excitations upon adding the extra matter fields including current and scalar condensate. Our study reveals some generic properties of boundary information measures dual to EWCS, e.g., entanglement of purification, logarithmic negativity and reflected entropy. Finally, we discuss how these results are consistent with the behavior of other correlation measures including the holographic mutual information. 
\end{abstract}
\end{titlepage}

\newpage

\tableofcontents
\noindent
\hrulefill

\onehalfspacing

\section{Introduction}\label{intro}

The gauge/gravity duality has stimulated a wide variety of recent efforts investigating the entanglement properties of quantum field theories (QFTs) and also the connection between entanglement and geometry, \textit{e.g.}, see \cite{Rangamani:2016dms,Nishioka:2018khk}. In this context, certain entanglement measures of the boundary QFT can be related to the geometric quantities that live in the bulk spacetime. The best studied example is the Ryu-Takayanagi (RT) prescription, which provides a holographic realization of the entanglement entropy corresponding to a spatial boundary subregion $A$\cite{hep-th/0603001}. In this case, the holographic entanglement entropy (HEE) can be computed as
\begin{eqnarray}\label{hee}
S_A={\rm min}\frac{{\rm area}(\Gamma_A)}{4G_N},
\end{eqnarray}
where $\Gamma_A$ is a bulk minimal hypersurface which ends on the boundary of $A$. It is important to highlight that entanglement entropy (EE) is a unique measure of quantum entanglement when the global state is pure. However, for a mixed state EE measures both classical and quantum correlations and in order to isolate the quantum correlations we should consider other measures, \textit{e.g.}, entanglement of purification (EoP), logarithmic negativity and reflected entropy. Much of our analysis in this paper will focus on studying EoP in different holographic settings, so we proceed by reviewing its definition.

EoP is a measure of total (both quantum and classical) correlations which coincides with EE for pure states. In general, it is a difficult quantity to obtain and only recently numerical lattice
calculations have been developed for QFTs\cite{Bhattacharyya:2018sbw,Camargo:2020yfv}. In order to define this quantity, let us consider a bipartite system with Hilbert space equal to the direct product of two factors, \textit{i.e.}, $\mathcal{H}=\mathcal{H}_A\otimes \mathcal{H}_B$. Assuming that a mixed state on $\mathcal{H}$ is described by a density matrix $\rho_{AB}$, we can purify it into a pure state $|\psi\rangle$ by adding auxiliary degrees of freedom to $\mathcal{H}$ such that $\rho_{AB}={\rm tr}_{A'B'}|\psi\rangle\langle\psi|$. In this case the EoP is defined as\cite{Terhal:2002}
\bea
E_P(A, B)\equiv \min_{|\psi\rangle} S(\rho_{A\cup A'}),
\eea
where $\rho_{A\cup A'}={\rm tr}_{BB'}|\psi\rangle\langle\psi|$. There are several useful inequalities which the EoP satisfies generally, \textit{e.g.},
\bea\label{iequalities}
&I(A, B)\leq 2E_P(A, B)\leq 2\;{\rm min}\{S_A, S_B\},\nonumber\\
&I(A, B)+I(A, C)\leq 2E_P(A, B\cup C),
\eea
where $I(A, B)$ is the mutual information given as follows
\bea\label{HMI}
I(A, B)=S_A+S_B-S_{A\cup B}.
\eea

\begin{figure}
\begin{center}
\includegraphics[scale=1]{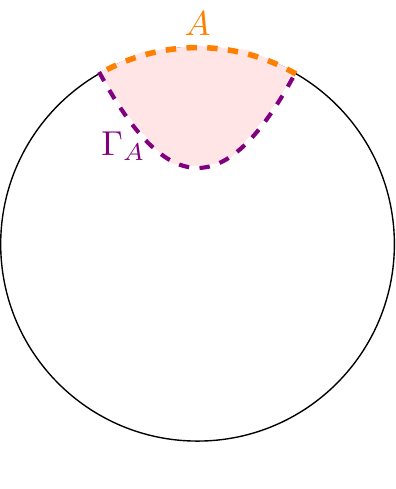}
\hspace*{1cm}
\includegraphics[scale=1]{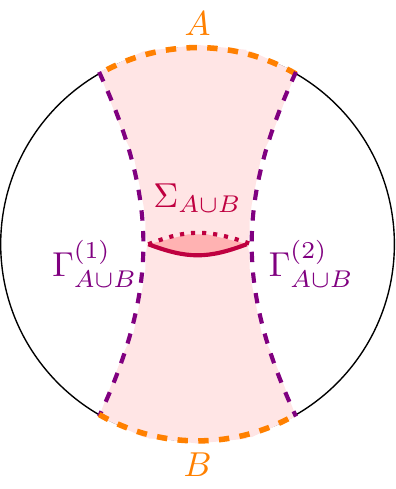}
\hspace*{1cm}
\includegraphics[scale=1]{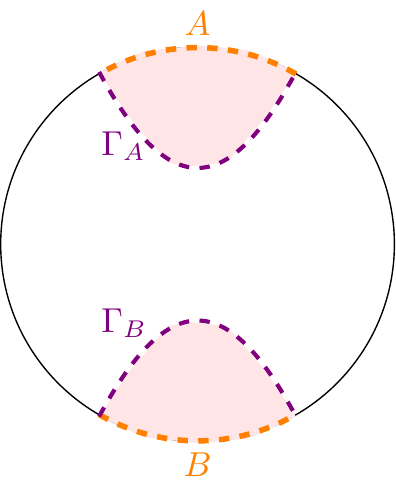}
\end{center}
\caption{\textit{Left}: The shaded region is the entanglement wedge corresponding to the boundary subregion $A$. \textit{Middle}: The case where the entanglement wedge is connected, \textit{i.e.}, $\Gamma_{A\cup B}=\Gamma_{A\cup B}^{(1)}\cup \Gamma_{A\cup B}^{(2)}$. In this case EWCS is proportional to the area of $\Sigma_{A\cup B}$. \textit{Right}: The case where the entanglement wedge is disconnected, \textit{i.e.}, $\Gamma_{A\cup B}=\Gamma_{A}\cup \Gamma_{B}$ and the corresponding EWCS vanishes.}
\label{fig:eopregions}
\end{figure}

Recently, there are many attempts to construct a holographic prescription for mixed state correlation measures  which have led to a remarkably rich and varied range of new insights, \textit{e.g.}, \cite{Takayanagi:2017knl,Nguyen:2017yqw,Dutta:2019gen,Kusuki:2019zsp}. A common feature of these studies is that all of the aforementioned correlation measures have a unique holographic counterpart which is the entanglement wedge cross section (EWCS). Before we proceed, let us recall the definition of EWCS in a simple holographic setup. Considering a spatial region A in the boundary field theory, the entanglement wedge is the bulk region corresponding to the reduced density matrix $\rho_A$ and whose boundary is $A\cup \Gamma_A$ (see the left panel in figure
\ref{fig:eopregions}). 
Note that for static geometries the entire configuration lies on a constant time slice. Similarly, this construction can be easily extended to more general cases where the boundary region is composed of multiple subregions. In particular, when the boundary region is union  of two disjoint subregions $A$ and $B$ the boundary of the entanglement wedge is $A\cup B \cup \Gamma_{A\cup B}$. In this case, keeping the geometry of $A$ and $B$ fixed while their separation
varies, the entanglement wedge has a phase transition due to the
competition between two different configurations for the corresponding RT hypersurface $\Gamma_{A\cup B}$ (see the middle and right panels in figure \ref{fig:eopregions}). For small separations the connected configuration is favored and we have $\Gamma_{A\cup B}=\Gamma_{A\cup B}^{(1)}\cup \Gamma_{A\cup B}^{(2)}$. On the other hand, for large separations the disconnected configuration is favored, and hence $\Gamma_{A\cup B}=\Gamma_{A}\cup \Gamma_{B}$. Now for connected configuration the EWCS is defined to be the minimal cross-sectional area of the entanglement wedge, \textit{i.e.}, 
\begin{eqnarray}\label{ew}
E_W(A, B)={\rm min}\frac{{\rm area}(\Sigma_{A\cup B})}{4G_N}.
\end{eqnarray}
Note that for disconnected configuration $\Sigma_{A\cup B}$ becomes empty and the corresponding $E_W$ vanishes.
In \cite{Takayanagi:2017knl,Nguyen:2017yqw}, a holographic proposal has been developed to describe the EoP corresponding to a general state in the boundary theory as follows
\begin{eqnarray}\label{ewep}
E_P(A, B)=E_W(A, B).
\end{eqnarray}
Based on this conjecture, it was shown that, the resultant quantity reproduces all the desired properties of EoP, \textit{e.g.}, eqs. \eqref{iequalities}. As we have already mentioned above there exist other boundary measures which seem to be dual to EWCS. Other proposals that make a connection between $E_W$ and different boundary correlation measures can be summarized as follows\cite{Dutta:2019gen,Kusuki:2019zsp}
\begin{eqnarray}\label{ewsr}
E_W(A, B)=\frac{S_R(A, B)}{2}=\frac{\mathcal{E}(A, B)}{\chi_d},
\end{eqnarray}
where $S_R$ and $\mathcal{E}$ are reflected entropy and logarithmic negativity, respectively.\footnote{However, we must add that the connection between EWCS and logarithmic negativity was recently called into question by \cite{Dong:2021clv}. In particular,
the derivations of \cite{Kusuki:2019zsp} were shown to not apply for bulk solutions that break
the replica symmetry.} Here $\chi_d$ is a constant which depends on the dimension of the spacetime. Let us also mention that in \cite{Tamaoka:2018ned} another measure of correlations the so-called odd (entanglement) entropy introduced which can be extracted from EWCS using
\begin{eqnarray}\label{odd}
S_O(A, B)=E_W(A, B)+S(A\cup B).
\end{eqnarray}
The above proposals pass a variety of consistency checks which provide important
pieces of evidence for finding holographic duals of EWCS. See \cite{Hirai:2018jwy,BabaeiVelni:2019pkw,Jokela:2019ebz,Umemoto:2019jlz,Akers:2019gcv,Amrahi:2020jqg,Chakrabortty:2020ptb,Saha:2021kwq} for various studies of general properties of EWCS and its holographic counterparts in static geometries. Further, the nonequilibrium evolution of EWCS for various quench protocols has been considered in \cite{1810.00420,1907.06646,2001.05501,BabaeiVelni:2020wfl,Moosa:2020vcs,Boruch:2020wbe}. Related investigations attempting to better understand the corresponding measures from the perspective of the boundary field theory have also appeared in \cite{1909.06790,Mollabashi:2020ifv,Bueno:2020fle,Camargo:2021aiq,Wen:2021qgx}. 

This paper presents another step in this research program, in which we investigate
the behavior of EWCS for different bulk geometries dual to boundary excited states. Indeed, developing a proper understanding of EWCS on the gravity side is essential to properly test the various holographic proposals noted above in eqs. \eqref{ewep}, \eqref{ewsr} and \eqref{odd}. In particular, we will consider states described by purely gravitational excitations in the bulk where the stress tensor is the only operator that has a nonvanishing expectation value. In this case, we will determine the leading corrections to $E_W$ and other correlation measures including HEE and HMI. We also consider more generic perturbations away from the vacuum, when other operators acquire nontrivial expectation values. Although for a generic excited state our analysis requires numerical treatment, we present some analytic results for the variation of EWCS in
the specific regimes of the parameter space. Let us recall that similar studies as we consider here were also considered in \cite{Bhattacharya:2012mi,Blanco:2013joa,Astaneh:2013gp,Allahbakhshi:2013rda,Wong:2013gua,Sheikh-Jabbari:2016znt,Bhattacharya:2019zkb,Saha:2020fon} to obtain insights into the behavior of HEE for excited states. We will elaborate more on these issues in the discussion section.


The rest of the paper is organized as follows: In section \ref{adsbb}, we briefly review the related analysis on EWCS for AdS black brane geometries including both thermal and charged excitations. In section \ref{adsplane}, we consider a specific class of anisotropic boundary excited states which are dual to AdS plane wave geometries and study the properties of EWCS, where we present both numerical and analytic results. Next, in section \ref{MomentumRelax} we study this quantity for momentum relaxation geometries where in the dual description additional scalar operators are excited. In section \ref{general}, we extend our studies to general boundary excited states which are dual to asymptotically AdS geometries. Finally, we briefly discuss our results and indicate some possible future directions, in section \ref{diss}.

\section{AdS Black Brane Geometries}\label{adsbb}
In this section we review the holographic information measures for AdS black brane geometries. In particular, we consider a $d$-dimensional boundary field theory with nonzero temperature and density dual to a charged AdS black brane geometry with a metric given by 
\begin{eqnarray}\label{metricadsbb}
ds^2=\frac{R^2}{r^2}\left(-f(r)dt^2+\sum_{i=1}^{d-1}dx_i^2+\frac{dr^2}{f(r)}\right),\hspace*{1.5cm}f(r)=1-\left(1+q^2r_h^2\right)\frac{r^d}{r_h^d}+q^2\frac{r^{2(d-1)}}{r_h^{2(d-2)}},
\end{eqnarray}
where $R$ and $r_h$ denote the AdS and horizon radius, respectively. Also $q$ is proportional to the charge density and related to the boundary chemical potential $\mu$, \textit{i.e.}, $q=\sqrt{\frac{d-2}{2d-2}}\mu$. From this metric, one obtains that the temperature and energy density of the equilibrium state are given by
\begin{eqnarray}\label{tempbb}
T=\frac{d}{4\pi r_h}\left(1-\frac{d-2}{d}q^2r_h^2\right),\hspace*{1.5cm}\mathcal{E}=\left(1+q^2r_h^2\right)\varepsilon,
\end{eqnarray}
where $\varepsilon=\frac{(d-1)R^{d-1}}{16\pi G_N}\frac{1}{r_h^d}$. Notice that the expectation value of the energy-momentum tensor dual to this geometry takes the form of that for an ideal fluid, \textit{i.e.},
\begin{eqnarray}
\langle T_{\mu\nu}\rangle={\rm diag}\left(1, \frac{1}{d-1},\cdots, \frac{1}{d-1}\right)\mathcal{E}.
\end{eqnarray}
In the extremal limit where the temperature vanishes we have $q^2r_h^2=\frac{d}{(d-2)}$ and the blackening factor has a double zero at $r=r_h$.

In order to find the HEE, in the following, we focus on the case of a strip entangling region lies on a constant time slice, \textit{i.e.},
\begin{eqnarray}\label{strip}
-\frac{\ell}{2}\leq x_1\equiv x\leq \frac{\ell}{2}\hspace*{2cm}0\leq x_2, \cdots, x_{d-1}\leq L
\end{eqnarray}
where $\ell\ll L$. In this case assuming that the minimal hypersurface has a translation invariance, its profile will be
determined by an embedding $x(r)$ and then using eqs. \eqref{hee} and \eqref{metricadsbb} the entropy functional is computed as 
\begin{eqnarray}\label{heefuncstrip}
S=\frac{R^{d-1}L^{d-2}}{4G_N}\int \frac{dr}{r^{d-1}}\sqrt{{x'}^2+\frac{1}{f(r)}},
\end{eqnarray}
where the prime indicates derivative with respect to $r$. The profile for the minimal hypersurface is then obtained by minimizing the above functional. Since there is no explicit $x(r)$ dependence, the corresponding momentum is a conserved quantity and hence the equation determining the profile simplifies to
\begin{eqnarray}\label{xp}
x'=\frac{1}{\sqrt{\left(\left(\frac{r_t}{r}\right)^{2d-2}-1\right)f(r)}},
\end{eqnarray}
where $r_t$ denotes the location of the turning point of $\Gamma_A$. Further, using the above expression the width of the entangling region can be written as follows
\begin{eqnarray}\label{Xadsbb}
\ell=2\int_0^{r_t} dr\frac{1}{\sqrt{\left(\frac{r_t^{2(d-1)}}{r^{2(d-1)}}-1\right)f(r)}}.
\end{eqnarray}
Also plugging eq. \eqref{xp} back into eq. \eqref{heefuncstrip} the HEE reads  
\begin{eqnarray}\label{heestat}
S=\frac{R^{d-1}L^{d-2}}{2G_N}\int_\epsilon^{r_t} \frac{dr}{r^{d-1}}\frac{1}{\sqrt{\left(1-\frac{r^{2(d-1)}}{r_t^{2(d-1)}}\right)f(r)}}.
\end{eqnarray}
On the other hand using holographic prescription we can find the HMI and EoP corresponding to a certain combined boundary region $A\cup B$. In the following, we consider a symmetric configuration consisting of two disjoint strips with equal width $\ell$ separated by $h$. Note that these measures are nontrivial only for connected configurations, \textit{i.e.}, $S_{A\cup B}=S(2\ell+h)+S(h)$, and vanish for disconnected ones, \textit{i.e.}, $S_{A\cup B}=2S(\ell)$, when $\Sigma_{A\cup B}$ becomes empty. Hence throughout the following, we will assume that the connected configuration is favored. In this case due to a reflection symmetry about $x=0$, $\Sigma_{A\cup B}$ runs along the radial direction and connects the corresponding turning points of $\Gamma_h$ and $\Gamma_{2\ell+h}$. Note that, keeping $\ell$ fixed while $h$ increases, the EWCS has a discontinuous  phase transition such that $\Sigma_{A\cup B}$ becomes empty when the two regions are distant enough \cite{Takayanagi:2017knl}. This behavior is due to the competition between two different configurations for the entanglement wedge. 
Assuming $\ell\gg h$, the connected configuration has the minimal area and we obtain a non-zero EWCS. Using eq. \eqref{ew} the corresponding functional becomes
\begin{eqnarray}\label{ewbb}
E_W=\frac{R^{d-1}L^{d-2}}{4G_N}\int_{r_d}^{r_u}\frac{dr}{r^{d-1}\sqrt{f}},
\end{eqnarray}
where $r_d$ ($r_u$) denotes the corresponding turning point of $\Gamma_h$ ($\Gamma_{2\ell+h}$). Finding the explicit dependence of HEE and EWCS on boundary quantities such as temperature or chemical potential requires computing the above integrals, which is a rather intricate task. So in the following, we present two specific examples in which we evaluate the variation of different measures due to the perturbations around the vacuum. We will consider thermal and charged excitations separately for which the leading order corrections can be obtained explicitly. 
\subsection{Thermal Excitations}
In this case we consider  $(q=0, T\neq 0)$ limit where the excited state is a thermal state at zero charge density. Focusing on low temperature limit we have $\ell T\ll 1$. From eq. \eqref{tempbb} this constraint can be expressed in terms of the bulk parameters as $r_t\ll r_h$. In this case
$r_t$ is close to the boundary and eq. \eqref{Xadsbb} has the leading behavior
\bea
\ell=r_t\left(c+\frac{\sqrt{\pi}}{(d+1)}\frac{\Gamma\left(\frac{d}{d-1}\right)}{\Gamma\left(\frac{d+1}{2d-2}\right)}\left(\frac{r_t}{r_h}\right)^{d}\right),
\eea
where $c=\frac{2\sqrt{\pi}\Gamma \left(\frac{d}{2d-2}\right)}{\Gamma \left(\frac{1}{2d-2}\right)}>0$.
Inverting this equation, we can represent the turning point as a function of $\ell$
\bea\label{rtsmalltemp}
r_t=\frac{\ell}{c}\left(1-\frac{\sqrt{\pi}}{(d+1)c^{d+1}}\frac{\Gamma\left(\frac{d}{d-1}\right)}{\Gamma\left(\frac{d+1}{2d-2}\right)}\left(\frac{\ell}{r_h}\right)^{d}\right).
\eea
That is, increasing the temperature, the turning point of the extremal hypersurface decreases. In this limit, the leading order behavior of eq. \eqref{heestat} reduces to
\begin{eqnarray}\label{heebb}
S=\frac{R^{d-1}L^{d-2}}{2(d-2)G_N}\left(\frac{1}{\epsilon^{d-2}}-\frac{c}{2r_t^{d-2}}+\frac{\sqrt{\pi}(d-2)}{4}\frac{\Gamma\left(\frac{d}{d-1}\right)}{\Gamma\left(\frac{d+1}{2d-2}\right)}\frac{r_t^2}{r_h^d}\right).
\end{eqnarray}
In principle then, we can invert the above expression to write our result in terms of the boundary quantities $\ell$ and $\varepsilon$. Combining the above results, we obtain the first order correction to HEE as follows
\begin{eqnarray}\label{svarBB}
\Delta S\equiv S-S_{\rm AdS}=\tilde{c}L^{d-2}\ell^2\varepsilon,
\end{eqnarray}
where
\begin{eqnarray}\label{tildec}
\tilde{c}=\frac{2\pi^{3/2}}{(d^2-1)c^2}\frac{\Gamma\left(\frac{1}{d-1}\right)}{\Gamma\left(\frac{d+1}{2d-2}\right)}.
\end{eqnarray}
Notice that $S_{\rm AdS}$ is the vacuum contribution given by $S_{\rm AdS}=\frac{R^{d-1}L^{d-2}}{2(d-2)G_N}\left(\frac{1}{\epsilon^{d-2}}-\frac{c^{d-1}}{2\ell^{d-2}}\right)$. In what follows we will use this $\Delta S$ notation in several parts of this paper. Note that $\tilde{c}>0$ and hence thermal excitations increase the HEE. This behavior is due to the fact that the number of degrees of freedom grows as we excite the system. It is worth to mention that the above relation between the leading order variation of HEE and the expectation value of the stress tensor, known as the first law of entanglement, was first noted in the holographic calculations of \cite{Bhattacharya:2012mi} and further derived in \cite{Blanco:2013joa} using the positivity of the relative entropy (see also \cite{Allahbakhshi:2013rda,Wong:2013gua} for related studies). Indeed, in \cite{Blanco:2013joa} it was shown that the first order variation in the entanglement entropy for a spatial subregion equals with the first order variation in the expectation value of the modular Hamiltonian which is a complicated object that cannot be expressed as an integral of local operators. Note that, in cases where the modular Hamiltonian is explicitly known, it is given by an integral of the energy density over the interior of the entangling region weighted by a shape dependent profile. In addition, as shown in \cite{Nozaki:2013vta} the first law of entanglement captures the same information as the linearized Einstein's equations.

Further, these results allow us to find the variation of other correlation measures. For example, considering a boundary configuration consisting of two disjoint strips with equal width $\ell$ separated by $h$, we can obtain the variation of HMI
\begin{eqnarray}\label{dIadsBB}
\Delta I\equiv I-I_{\rm AdS}=- 2\tilde{c}L^{d-2}\left(\ell+h\right)^2\varepsilon,
\end{eqnarray}
where $I_{\rm AdS}$ is the vacuum contribution given by $I_{\rm AdS}=\frac{R^{d-1}L^{d-2}c^{d-1}}{2(d-2)G_N}\left(\frac{1}{2\left(2\ell+h\right)^{d-2}}+\frac{1}{2h^{d-2}}-\frac{1}{\ell^{d-2}}\right)$. The minus sign shows that the thermal excitations decrease the HMI and hence reduce the total correlation between the subregions.
 
Next we examine the variation of EoP due to the increase in energy at low temperature using eq. \eqref{ewbb}. At this limit, we can expand $E_W$ for large $r_h$, which yields \cite{BabaeiVelni:2019pkw}
\begin{eqnarray}
E_W=\frac{R^{d-1}L^{d-2}}{4(d-2)G_N}\left(\frac{1}{r_d^{d-2}}-\frac{1}{r_u^{d-2}}\right)+\frac{R^{d-1}L^{d-2}}{16G_N}\frac{r_u^2-r_d^2}{r_h^d}.
\end{eqnarray}
Now, we use eq. \eqref{rtsmalltemp} to express $r_d$ and $r_u$ in terms of $\ell$ and $h$. As a result, the variation of EoP becomes
\begin{eqnarray}\label{ewvarBB}
\Delta E_W\equiv E_W-E_{W\rm AdS}=\mathcal{C}L^{d-2}\ell(\ell+h)\varepsilon,
\end{eqnarray}
where 
\begin{eqnarray}\label{mathcalc}
\mathcal{C}=\frac{4\pi}{(d-1)c^2}\left(1-\frac{4\sqrt{\pi}}{(d+1)c}\frac{\Gamma\left(\frac{d}{d-1}\right)}{\Gamma\left(\frac{d+1}{2d-2}\right)}\right),
\end{eqnarray}
and $E_{W\rm AdS}$ is the vacuum contribution given by $E_{W\rm AdS}=\frac{R^{d-1}L^{d-2}c^{d-2}}{4(d-2)G_N}\left(\frac{1}{h^{d-2}}-\frac{1}{\left(2\ell+h\right)^{d-2}}\right)$. Notice that $\mathcal{C}<0$ and hence the finite temperature corrections decrease the EoP. Regarding the EoP as a measure of total correlation between the two subregions, we see that thermal excitations promote disentangling between them. Further, comparing eq. \eqref{ewvarBB} with eq. \eqref{svarBB} we see that in the case of two adjacent intervals, \textit{i.e.}, $h\ll \ell$, the variation of EoP is proportional to the variation of HEE up to a negative constant. At this point, let us recall that in $d=2$ the variation of these measures in the same limit  takes the following simple form \cite{BabaeiVelni:2019pkw}
\begin{eqnarray}\label{dsdewd2}
\Delta S=\frac{\pi}{3}\ell^2 \varepsilon+\mathcal{O}\left(\frac{h}{\ell}\right),\hspace*{2cm}\Delta E_W=-\frac{\pi}{3}\ell^2 \varepsilon+\mathcal{O}\left(\frac{h}{\ell}\right),
\end{eqnarray}
which shows that at leading order $\Delta E_W=-\Delta S$ and hence the measures change with the same rate (in this case we have $|\mathcal{C}|=\tilde{c}=\frac{\pi}{3}$). On the other hand, in higher dimensions $|\mathcal{C}|<\tilde{c}$ and thus the variation of HEE becomes more pronounced. 

Before closing this section, let us comment further on the computation of the first order corrections to the holographic correlation measures, \textit{e.g.}, HEE and EoP, under the metric perturbation. In the absence of perturbations, we have a pure AdS geometry and the HEE for a spatial subregion $A$ is given in terms of the area of a minimal hypersurface $\Gamma_A$. Now, consider an arbitrary small perturbation to the AdS geometry which is dual to small deviation of the boundary vacuum state. We expect that both the area functional and the shape of the extremal hypersurface are modified and the corresponding HEE changes. More explicitly, we have $\Gamma_A\rightarrow \Gamma_A+\delta\Gamma_A$ and $S_A\rightarrow S_A+\delta S_A$. However, considering the first order corrections to HEE, one can show that any change in the profile of the minimal hypersurface will not contribute to $\delta S_A$. Hence we can evaluate the linear change in the HEE by evaluating the area functional on  $\Gamma_A$ in the perturbed background \cite{Nozaki:2013vta}. Note that in this case the boundary subregion $A$ is fixed and thus the boundary condition for the minimal hypersurface does not change.

Turning now to the computation of the first order correction to EoP, we should find the minimal cross-sectional area of the entanglement wedge, \textit{i.e.}, $\Sigma_{A\cup B}$, in the perturbed geometry. Recall that by definition, the entanglement wedge is bounded by $A\cup B \cup \Gamma_{A\cup B}$ and hence any change in the position of the RT hypersurface, modifies the profile of $\Sigma_{A\cup B}$. This means that the boundary condition for $\Sigma_{A\cup B}$ changes under metric perturbation, while keeping $A$ and $B$ fixed. Hence we expect that the variation of $\Gamma_{A\cup B}$ plays a central role in evaluating the correction to EoP even at leading order. As an example, consider the symmetric configuration consisting of two disjoint strips with equal width. In this case due to the reflection symmetry $\Sigma_{A\cup B}$ lies entirely on $x=0$ slice both for pure AdS geometry and for any homogeneous excitations around it.\footnote{Here by homogeneous excitations we mean excitations being independent of $x$, so that the profile of EWCS does not change.} On the other hand, because the turning points of the corresponding RT hypersurfaces, \textit{i.e.}, $\Gamma_{h}$ and $\Gamma_{2\ell+h}$, are modified the limits of integration in eq. \eqref{ewbb} change which give a nontrivial contribution to $\Delta E_W$. Further, we note that in the present example, turning on the temperature, the RT hypersurfaces move away from the horizon to smaller values of $r$. In the following, we will see that similar results apply for other types of excitations.

\subsection{Charged Excitations}
In this case we consider a charged black brane which is dual to a finite temperature boundary state with a nonzero chemical potential for some conserved charge. We assume that the chemical potential is of the same order as temperature, \textit{i.e}, $\mu \sim T$, and of course, as in the
previous case, we assume $ \ell T\ll 1$. Again, these constraints can be expressed in terms of the bulk parameters as $qr_h\sim \mathcal{O}(1)$ and $r_t\ll r_h$. Thus, $r_t$ is close to the boundary and eq. \eqref{Xadsbb} yields
\bea\label{rtsmalltempq}
r_t=\frac{\ell}{c}\left(1-\frac{\sqrt{\pi}}{(d+1)c^{d+1}}\frac{\Gamma\left(\frac{d}{d-1}\right)}{\Gamma\left(\frac{d+1}{2d-2}\right)}\left(\frac{\ell}{r_h}\right)^{d}\left(1+q^2r_h^2\right)+\frac{\sqrt{\pi}d}{(2d-1)c^{2d-1}}\frac{\Gamma\left(\frac{d}{2d-2}\right)}{\Gamma\left(\frac{1}{2d-2}\right)}\left(\frac{\ell}{r_h}\right)^{2d-2}q^2r_h^2\right).
\eea
We see once again (as in the neutral case) that at leading order the turning point is a decreasing function of the excitation parameter. Substituting the above expression into eq. \eqref{heestat}, we obtain
\begin{eqnarray}\label{deltasq}
\Delta S(q)=\left(1+q^2r_h^2\right)\Delta S-\frac{R^{d-1}L^{d-2}}{4G_N}\frac{\sqrt{\pi}(d-1)}{(2d-1)c^d}\frac{\Gamma\left(\frac{d}{2d-2}\right)}{\Gamma\left(\frac{1}{2d-2}\right)}q^2r_h^2\frac{\ell^d}{r_h^{2d-2}},
\end{eqnarray}
where $\Delta S$ is defined in eq. \eqref{svarBB}. Note that at leading order the charged excitations increase the HEE, while higher order correction is negative. A similar calculation for EoP functional eq. \eqref{ewbb} shows that
\begin{eqnarray}
E_W=\frac{R^{d-1}L^{d-2}}{4G_N}\left(\frac{1}{(d-2)r_d^{d-2}}-\frac{1}{(d-2)r_u^{d-2}}+(1+q^2r_h^2)\frac{r_u^2-r_d^2}{4r_h^d}-\frac{q^2r_h^2}{2d}\frac{r_u^d-r_d^d}{r_h^{2d-2}}\right).
\end{eqnarray}
Hence using eq. \eqref{rtsmalltempq} the variation of EoP becomes
\begin{eqnarray}\label{ewbbext}
\Delta E_W(q)=\left(1+q^2r_h^2\right)\Delta E_W+\frac{R^{d-1}L^{d-2}}{4G_N}\mathcal{C}_qq^2r_h^2\frac{(2\ell+h)^d-h^d}{r_h^{2d-2}},
\end{eqnarray}
where $\Delta E_W$ is given by eq. \eqref{ewvarBB} and $\mathcal{C}_q$ is a constant factor which can be written as
\begin{eqnarray}
\mathcal{C}_q=\frac{\sqrt{\pi}d}{(2d-1)c^{d+1}}\frac{\Gamma\left(\frac{d}{2d-2}\right)}{\Gamma\left(\frac{1}{2d-2}\right)}-\frac{1}{2dc^d}.
\end{eqnarray}
Therefore, the charged excitations decrease the EoP at leading order and hence reduce the total correlation between the two subregions. Note that $\mathcal{C}_q>0$ and thus the subleading term is positive. Further, using eq. \eqref{deltasq} we see that the corresponding HMI decreases at leading order. This is perfectly consistent with eq. \eqref{ewbbext} since both HMI and EoP are measures of total correlation between the subregions.

\section{AdS Plane Wave Geometries}\label{adsplane}
In this section, we would like to generalize the previous analysis to determine corrections to the EoP in a class of boundary excited states which are dual to the AdS plane wave geometries. These gravitational backgrounds are dual to anisotropically excited states in a QFT with a constant energy flux. The bulk geometry will be an asymptotically AdS$_{d+1}$ black brane with a metric given by \cite{Narayan:2012hk}
\begin{equation}\label{m1}
ds^2=\frac{R^2}{r^2}\left(-2dx^+dx^-+\sum_{i=1}^{d-2}dx_i^2+dr^2\right)+R^2\mathcal{T}r^{d-2}(dx^+)^2.
\end{equation}
Here $\mathcal{T}$ is proportional to the boundary energy density, \textit{i.e.}, $\mathcal{T}\propto T_{++}$ and based on the null energy condition, the range $\mathcal{T}>0$ is allowed. The $x^+$ and $x^-$ coordinates are light-like defined as 
\begin{equation}
x^{\pm}=\frac{t\pm x_{d-1}}{\sqrt{2}}.
\end{equation}
As shown in \cite{Singh:2010zs,Singh:2012un} the metric \eqref{m1} arises from the boosted AdS black brane in the limit of infinite boost. Let us recall that the boosted AdS black brane geometries, also known as the regularized AdS plane waves, are dual to a uniformly boosted strongly coupled large $N$ thermal state. The corresponding metric is given by \cite{Blanco:2013joa}
\begin{eqnarray}\label{BBB}
ds^2=\frac{R^2}{r^2}\left(-dt^2+dx_{d-1}^2+\gamma^2\frac{r^d}{r_h^d}\left(dt+v\,dx_{d-1}\right)^2+\sum_{i=1}^{d-2}dx_i^2+\frac{dr^2}{f(r)}\right),\hspace*{1cm}f(r)=1-\frac{r^d}{r_h^d},
\end{eqnarray}
where $v$ is the velocity and $\gamma^{-2}=1-v^2$. This geometry is dual to a thermal plasma which is boosted along the $x_{d-1}$ direction with the following expressions for temperature and energy density
\begin{eqnarray}
T=\frac{d}{4\pi\gamma r_h},\hspace*{2cm}\mathcal{E}=\left(1+\frac{d}{d-1}\gamma^2v^2\right)\varepsilon.
\end{eqnarray}
Note that defining $\mathcal{T}=\frac{(1+v)^2}{2(1-v^2)r_h^d}$ and considering $v\rightarrow 1$ while $r_h\rightarrow \infty$ one can show that eq. \eqref{BBB} reduces to eq. \eqref{m1}. It is important to mention that the dual states described by purely gravitational excitations in the bulk because the stress tensor is the only operator that has a nonvanishing expectation value.

In the following, we first study the behavior of the correlation measures in a boosted AdS black brane and then we examine their variation in the limit of infinite boost. 
Note that, due to the boosting along a certain direction the rotational symmetry in the boundary theory is broken and the corresponding HEE shows a number of interesting features \cite{Bhatta:2019eog,Maulik:2020tzm}.  It is worth to mention
that the HEE in AdS plane wave geometry has been studied in \cite{Narayan:2012ks}. Also, the behavior and phase  transitions of HMI in this background has been analyzed in \cite{Mukherjee:2014gia}.

\subsection{EWCS in Boosted AdS Black Brane}\label{sec:BBB}
In this section, we apply holographic prescription to find the EoP using eq. \eqref{ew} for configurations consisting of thin long strips. In order to investigate the behavior of EoP, we consider two different types of boundary subregions: (a) the width of the entangling region is orthogonal to the direction of boost, \textit{i.e.}, $x_{d-1}$, called case A, or (b) the width of the entangling region is along the $x_{d-1}$-direction, called case B.

\subsection*{Case A}

Considering the strip entangling region to be along $x_1$-direction, we can parameterize the minimal hypersurface by $x_1=x(r)$. Then the HEE is computed as 
\begin{eqnarray}
S=\frac{R^{d-1}L^{d-2}}{4G_N}\int \frac{dr}{r^{d-1}}\sqrt{1+v^2\gamma^2\frac{r^d}{r_h^d}}\sqrt{{x'}^2+\frac{1}{f}}.
\end{eqnarray}
By minimizing this entropy functional, we can determine the profile of $x(r)$. Once again, we have a conserved quantity and hence finding the minimal hypersurface is a straightforward exercise. Plugging the resultant profile in the above functional, we are left with
\begin{eqnarray}\label{heebbb}
S=\frac{R^{d-1}L^{d-2}r_t^{d-1}}{2G_N}\int_0^{r_t}\frac{dr}{r^{2d-2}}\frac{1+v^2\gamma^2 \frac{r^d}{r_h^d}}{\sqrt{f}}\frac{1}{\sqrt{\left(\frac{r_t}{r}\right)^{2(d-1)}+v^2\gamma^2\frac{r_t^d}{r_h^d}\left(\left(\frac{r_t}{r}\right)^{d-2}-1\right)-1}},
\end{eqnarray}
where the relation between $\ell$ and $r_t$ is given by
\begin{eqnarray}\label{Xboost}
\ell=2\int_0^{r_t}\frac{dr}{\sqrt{f}\sqrt{\left(\frac{r_t}{r}\right)^{2(d-1)}\frac{r_h^d+v^2\gamma^2 r^d}{r_h^d+v^2\gamma^2 r_t^d}-1}}.
\end{eqnarray}
Next, we can also compute EoP in this symmetric configuration for a boundary state dual to eq. \eqref{BBB}. In this case the corresponding expression can be written as
\begin{eqnarray}\label{ewbbb}
E_W=\frac{R^{d-1}L^{d-2}}{4G_N}\int_{r_d}^{r_u}\frac{dr}{r^{d-1}\sqrt{f}}\sqrt{1+v^2\gamma^2 \frac{r^d}{r_h^d}}.
\end{eqnarray}
Now we are equipped with all we need to study the desired correlation measures in this background. Before examining the full dependence of different correlation measures on boost parameters, we would like to study the leading order variation of them. First, we expand eq. \eqref{Xboost} in the limit
$r_t\ll r_h$ to find the leading corrections to $r_t$ compared to its static value
\begin{equation}\label{rtbbb}
r_t=r_t^{(0)}+\frac{\sqrt{\pi}}{(d-1)c^{d+2}}\left(\frac{\Gamma(\frac{d}{2d-2})}{\Gamma(\frac{1}{2d-2})}-\frac{\Gamma(\frac{d}{d-1})}{\Gamma(\frac{d+1}{2d-2})}\right)v^2\gamma^2\frac{\ell^{d+1}}{r_h^d},
\end{equation}
where $r_t^{(0)}$ is the turning point for $v=0$ given by eq. \eqref{rtsmalltemp}. 
Using the above expression and eq. \eqref{heebbb}, the HEE can be written in terms of the following expansion
\begin{equation}
S=S^{(0)}+\frac{2 \pi ^{3/2} L^{d-2} \Gamma \left(\frac{d}{d-1}\right)}{c^2 (d-1) \Gamma \left(\frac{d+1}{2 d-2}\right)} v^2\gamma ^2  \ell^2\varepsilon,
\end{equation}
where again $S^{(0)}$ is the HEE for $v=0$, \textit{i.e.}, eq. \eqref{heebb}. In comparing the above expression with eq. \eqref{svarBB}, we see that  
\begin{eqnarray}\label{dsboost}
\Delta S'\equiv S-S^{(0)}=\left(1+\frac{d+1}{d-1}v^2\gamma^2\right)\Delta S.
\end{eqnarray}
Of course, this reproduces the result first found in \cite{Blanco:2013joa}. Clearly, the HEE increases after the boost transformation. Using this result the variation of HMI becomes
\begin{eqnarray}\label{dIboost}
\Delta I'=\left(1+\frac{d+1}{d-1}v^2\gamma^2\right)\Delta I.
\end{eqnarray}
Now we proceed to examine the behavior of EoP in low temperature limit. First, we expand eq. \eqref{ewbbb} in the limit $r_d\ll r_u \ll r_h$ to find the leading corrections to $E_W$. Next, we use eq. \eqref{rtbbb} for $r_d$ and $r_u$ separately to express the variation of EoP in terms of the boundary quantities as follows
\begin{align}
\Delta E_W'\equiv {E_W}-E_W^{(0)}=\frac{d+1}{d-1}L^{d-2} v^2\gamma ^2 \ell(h+\ell)\varepsilon.
\end{align}
We note again that using eq. \eqref{ewvarBB}, the above result can be written as follows 
\begin{eqnarray}
\Delta E_W'=\left(1+\frac{d+1}{d-1}v^2\gamma^2\right)\Delta E_W.
\end{eqnarray}
Comparing the above result with eqs. \eqref{dsboost} and \eqref{dIboost} shows that the boost transformation has the same effect on the correlation measures as expected. 

\begin{figure}
\begin{center}
\includegraphics[scale=0.59]{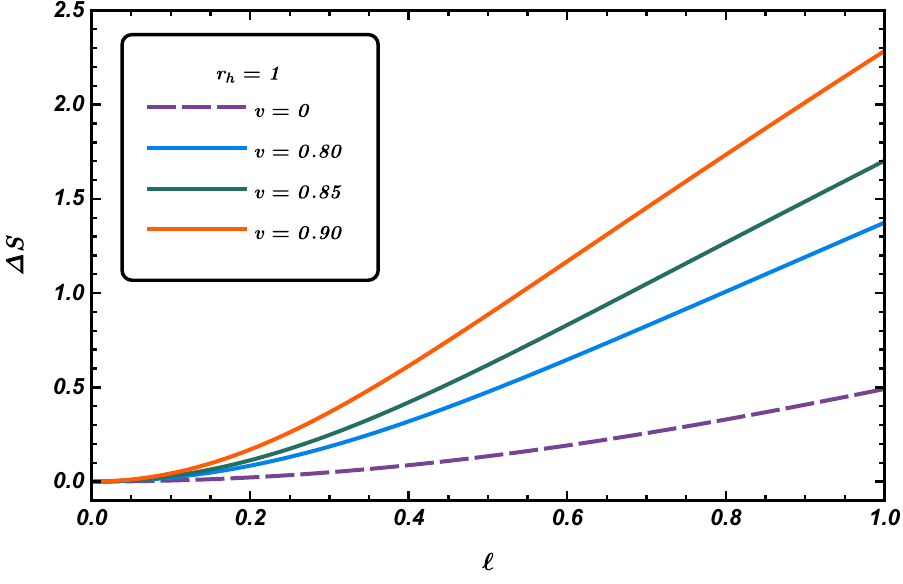}
\hspace*{0.04cm}
\includegraphics[scale=0.58]{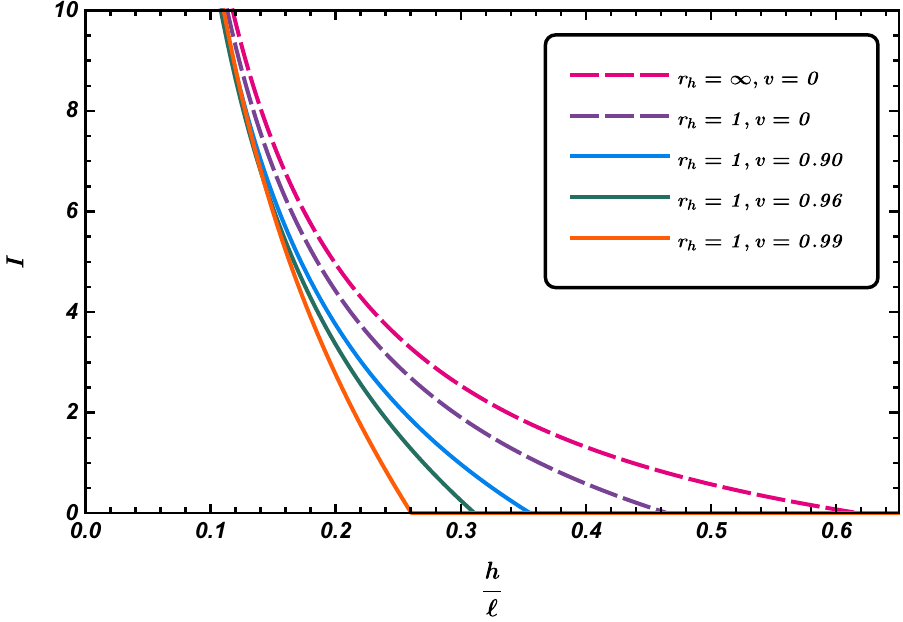}
\hspace*{0.04cm}
\includegraphics[scale=0.58]{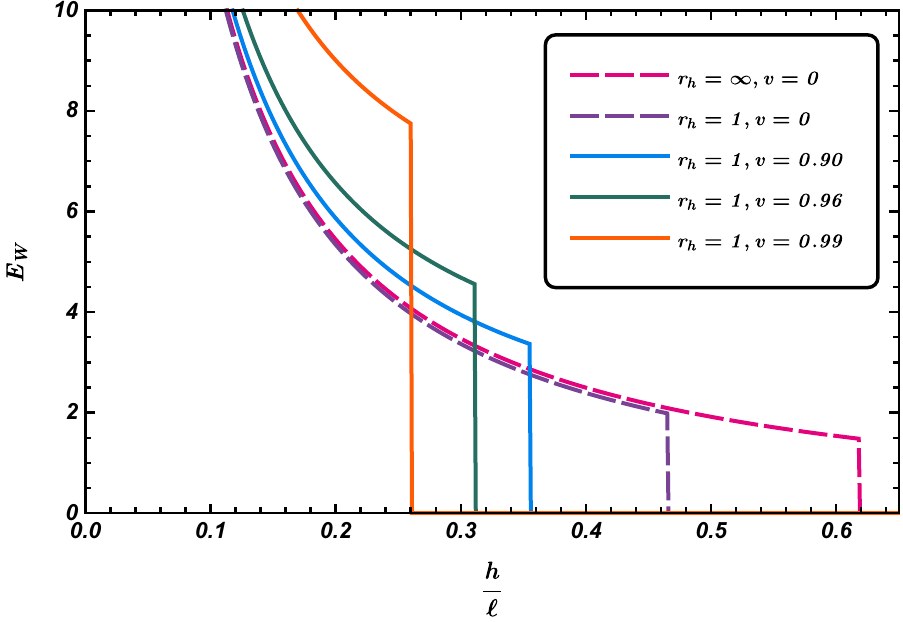}
\end{center}
\caption{Subtracted HEE (left), HMI (middle) and EWCS (right) as functions of the width and separation between subregions for different values of $r_h$ and $v$ in $d=3$.}
\label{fig:BBB3d}
\end{figure}

Now we turn our attention to numerically evaluating the correlation measures for different values of the parameters. We will mainly focus on three dimensional boundary theory, because the interesting qualitative features of the measures are independent of the dimensionality of the field theory. In figure \ref{fig:BBB3d} we show the subtracted HEE, the corresponding HMI and EoP for specific values of $h$ and $\ell$. For simplicity, we have rescaled the holographic measures, \textit{i.e.}, $\{S, I, E_W\}\rightarrow \frac{4G_N}{R^{d-1}L^{d-2}}\{S, I, E_W\}$. The left panel shows the dependence of HEE on the width of the entangling region for different values of temperature and velocity. Note that the HEE is regularized by subtracting the AdS contribution. The dashed violet curve corresponds to AdS black brane geometry. The middle panel demonstrates the HMI as a function of the dimensionless boundary quantity $h/\ell$. Here the dashed purple curve corresponds to pure AdS geometry.  In the right panel we show $E_W$ for the same values of the parameters. Based on this plots, we observe that although the HEE and EoP increase with the boost parameter, the HMI decreases with $v$. Also the phase transition of EoP happens at smaller separations between the two subregions comparing to $v=0$ case.
\subsection*{Case B}
\begin{figure}
\begin{center}
\includegraphics[scale=0.66]{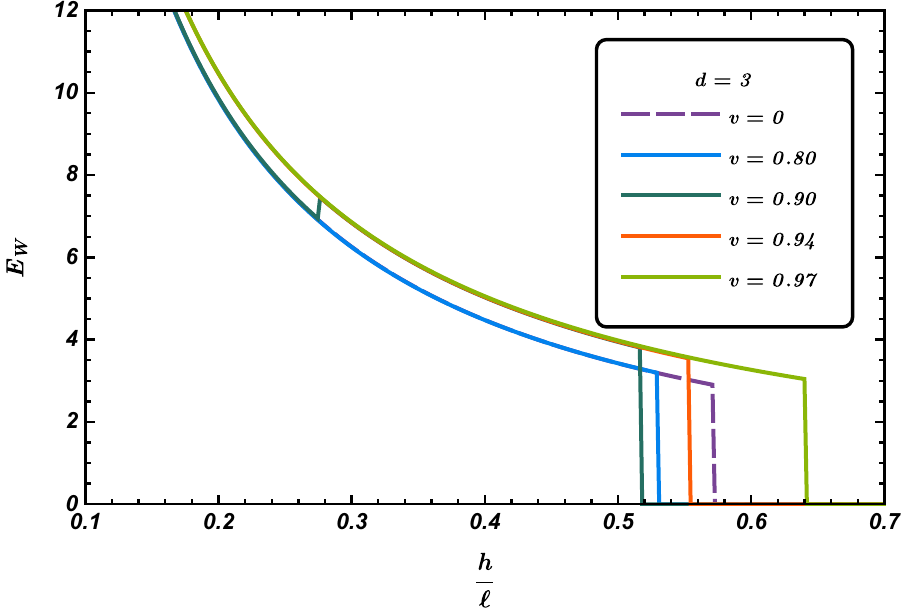}
\hspace*{1cm}
\includegraphics[scale=0.66]{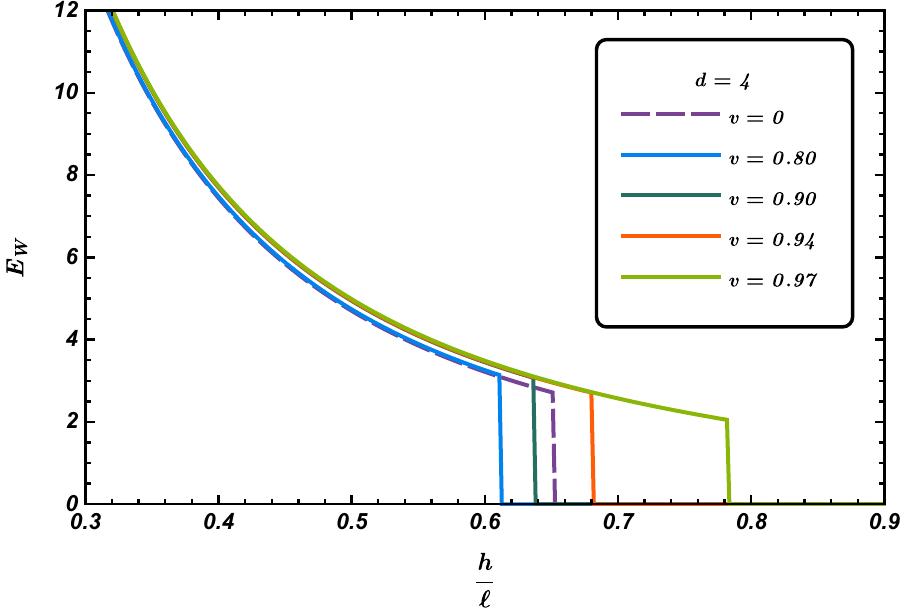}
\end{center}
\caption{EoP for $d=3$ (left) and $d=4$ (right) as functions of the width and separation between subregions with $r_h=1$ for different values of $v$.}
\label{fig:BBBcaseB}
\end{figure}

In this case the width of the strip entangling region is along $x_{d-1}$-direction and we can parameterize the minimal hypersurface by $x_{d-1}\equiv x(r)$. Given the metric in eq. \eqref{BBB} the corresponding entropy functional becomes
\begin{eqnarray}\label{heecaseb}
S=\frac{R^{d-1}L^{d-2}}{4G_N}\int \frac{dr}{r^{d-1}}\sqrt{\left(1+v^2\gamma^2 \frac{r^d}{r_h^d}\right){x'}^2+\frac{1}{f}}.
\end{eqnarray}
Applying the same steps as in our previous calculations, one can evaluate the leading corrections to the HEE at small temperature as
\begin{equation}
\Delta'' S=\left(1+\frac{2}{d-1}v^2\gamma ^2\right)\Delta S, 
\end{equation}
where $\Delta S$ was defined in eq. \eqref{svarBB}. We again use the symmetry of the configuration to  evaluate the EoP in this case. The corresponding area functional can be obtained by setting $x'=0$ in eq. \eqref{heecaseb}. Expanding the integrand  and evaluating the resultant expression we have
\begin{equation}
E_W=\frac{R^{d-1}L^{d-2}}{4G (d-2)}\left(\frac{1}{r_d^{d-2}}-\frac{1}{r_u^{d-2}}\right)+\frac{R^{d-1}L^{d-2}}{16 G r_h^d }\left(r_u^2-r_d^2\right),
\end{equation}
which can be re-express as follows
\begin{equation}
\Delta'' E_W=\left(1+\frac{2}{d-1}v^2\gamma ^2\right)\Delta E_W,
\end{equation}
where $\Delta E_W$ was defined in eq. \eqref{ewvarBB}. In obtaining the above result, we have used that at leading order
\begin{equation}
r_t=r_t^{(0)}-\frac{\sqrt{\pi } \gamma ^2 v^2 \ell^{d+1}}{(d-1)c^{d+2}r_h^d}\left(\frac{2 \Gamma \left(\frac{d}{d-1}\right)}{(d+1) \Gamma \left(\frac{d+1}{2 d-2}\right)}-\frac{\Gamma \left(\frac{d}{2 d-2}\right)}{\Gamma \left(\frac{1}{2 d-2}\right)}\right).
\end{equation}
To close this section, we present plots of EoP in various dimensions in fig. \ref{fig:BBBcaseB} for several values of $v$. Clearly the EoP is an increasing function of the boost parameter which is consistent with our  previous results. Note that the derivative of EoP for $v=0.9$ in the left panel has a discontinuity which is due to the appearance of a new disconnected configuration for the corresponding RT hypersurface. As discussed in \cite{Narayan:2012ks}, there is a disconnected hypersurface simply given by $x'=0$ such that for small regions the connected surface has the minimal area, while for large regions the RT surface changes
topology and the disconnected surface is favored. 
\subsection{EWCS in AdS Plane Wave: Case A}\label{APWcaseA}
\begin{figure}
\begin{center}
\includegraphics[scale=0.59]{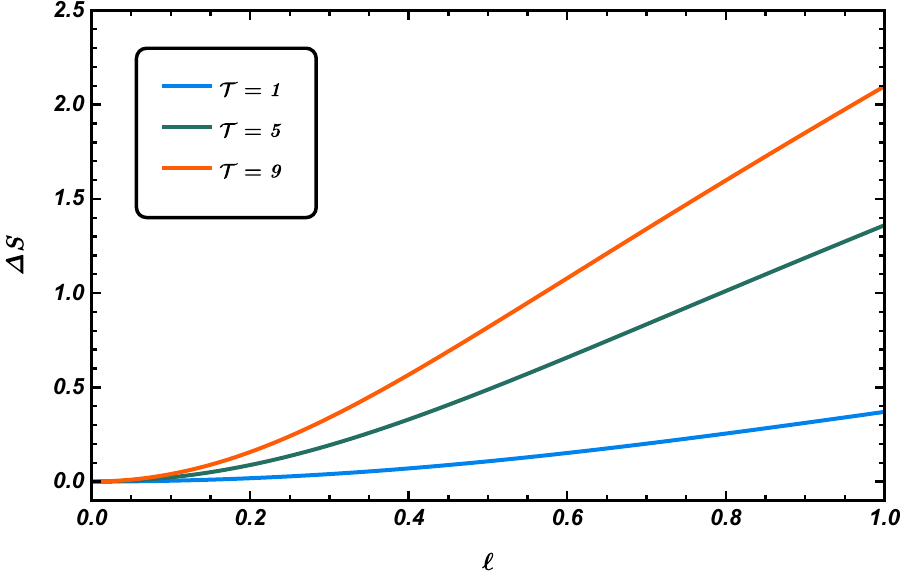}
\hspace*{0.04cm}
\includegraphics[scale=0.58]{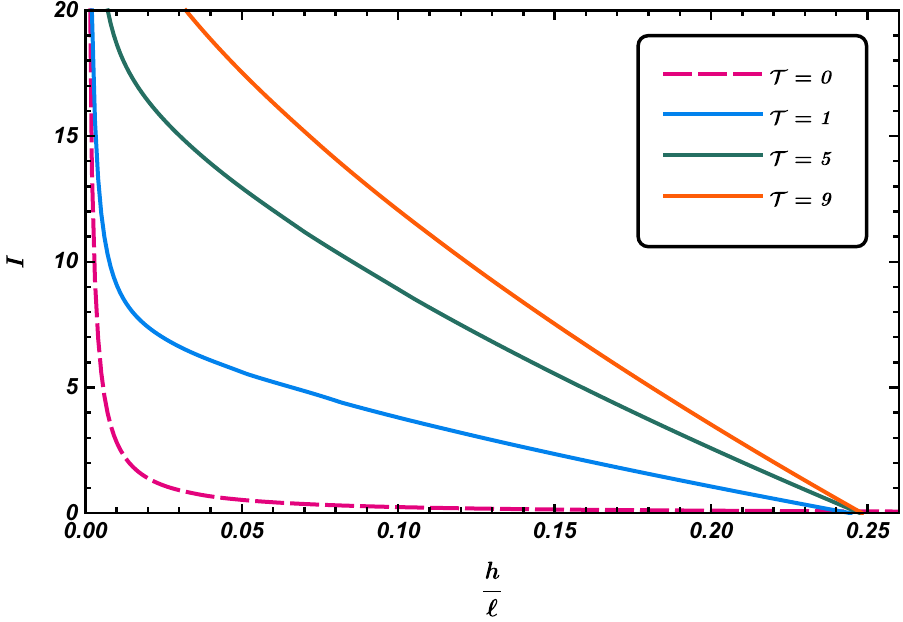}
\hspace*{0.04cm}
\includegraphics[scale=0.58]{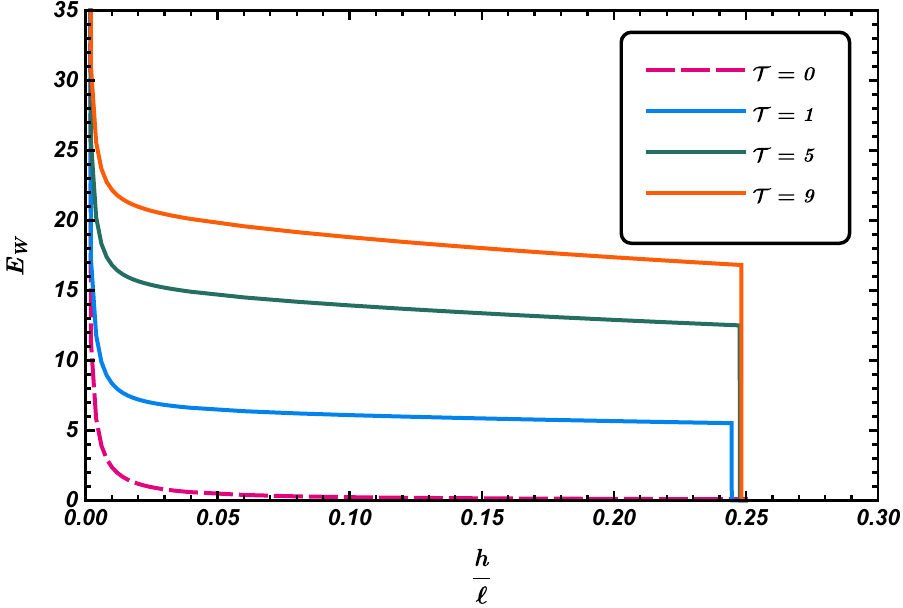}
\end{center}
\caption{Subtracted HEE (left), HMI (middle) and EoP (right) as functions of the width and separation between subregions for different values of $\mathcal{T}$ in $d=3$.}
\label{fig:APW3dcaseA}
\end{figure}

In this section, our goal is to calculate the variation of EoP in an AdS plane wave geometry when the width of the strip is along the $x_1$-direction. Recall that in this case the energy flow is parallel with the entangling region. Upon substituting the profile of the minimal hypersurface, \textit{i.e.}, $x_1\equiv x(r)$ into the metric \eqref{m1}, the HEE functional simplifies to
\begin{equation}
S=\frac{R^{d-1} L^{d-2}}{4G_N}\int \frac{dr}{r^{d-1}}\sqrt{(1+{x'}^2)(1+\frac{\mathcal{T} r^d}{2} )}.
\end{equation}
Minimizing the above expression yields the equation of motion for $x(r)$ which reads
\begin{equation}
{x'}^2=\frac{2+\mathcal{T} r_t^d}{(2+\mathcal{T} r^d)r_t^{2d-2}-(2+\mathcal{T} r_t^d)r^{2d-2}}r^{2d-2}.
\end{equation}
We note again that $r_t$ is the turning point of the minimal hypersurface which is fixed by setting $x'(r_t)=\infty$. Further, using the above result the width of the entangling region and the HEE can be written in terms of the following expressions
\begin{eqnarray}
\ell&=&2\sqrt{2+\mathcal{T} r_t^d}\int_{0}^{r_t} dr\;r^{d-1}\;\frac{1}{\sqrt{(2+\mathcal{T} r^d)r_t^{2d-2}-(2+\mathcal{T} r_t^d)r^{2d-2}}},\label{lAPW}\\
S&=&\frac{L^{d-2} R^{d-1}}{2G_N}\int_{\epsilon}^{r_t} \frac{dr}{r^{d-1}} \frac{(2+\mathcal{T}r^d)r_t^{d-1}}{\sqrt{2(2+\mathcal{T}r^d)r_t^{2d-2}-2(2+\mathcal{T}r_t^d)r^{2d-2}}}.\label{SAPW}
\end{eqnarray}
Note that using the above expressions one can easily show that the divergent part of HEE is the standard area law which does not depend on $\mathcal{T}$ as we expected, since the metric \eqref{m1} is asymptotically AdS. Also we can evaluate the EWCS for the same configuration of entangling regions that we considered before. Once again, since there is a reflection symmetry, we expect that $\Sigma$ lies entirely on $x=0$ slice. Thus the EoP can be written as follows
\begin{equation}\label{EWCSr}
E_W=\frac{R^{d-1}L^{d-2}}{4G_N}  \int_{r_d}^{r_u}\frac{dr}{r^{d-1}}\sqrt{1+\frac{\mathcal{T}}{2}r^{d}}.
\end{equation}
In figure \ref{fig:APW3dcaseA} we show the HEE, the corresponding HMI and EoP for specific values of $h, \ell$ and $\mathcal{T}$. The left panel shows the subtracted HEE as a function of $\ell$ for different values of $\mathcal{T}$. The middle panel demonstrates the HMI as a function of $h/\ell$ which has a continuous transition due to the competition between connected and disconnected configurations. Again, the dashed purple curve corresponds to AdS black brane geometry. In the right panel we show EoP for the same values of the parameters.  As is clear from the graphs, all three correlation measures are monotonically increasing function of the energy density. 

To better understand these behaviors, let us explore the scaling of the correlation measures in the specific regimes of the parameter space. A simple analysis shows that in $\mathcal{T}\ell^d \gg 1$ limit, we have $\ell \sim r_t$ and the finite part of the HEE takes the following form \cite{Narayan:2012ks}
\begin{equation}\label{ds}
S_f\sim \frac{R^{d-1}L^{d-2}}{G_{N}}  \sqrt{\mathcal{T}}\begin{cases}
\pm\ell^{2-\frac{d}{2}}&\qquad d\neq 4\\
\log(\ell \mathcal{T}^{1/4}) &\qquad d=4
\end{cases},
\end{equation}
where $+(-)$ sign is for $d < 4(> 4)$. Clearly, the finite part of HEE is always smaller than the thermal entropy and is a monotonically increasing function of $\mathcal{T}$ and $\ell$. The latter is due to the fact that the number of degrees of freedom grows as we excite the system.

\begin{figure}
\begin{center}
\includegraphics[scale=0.75]{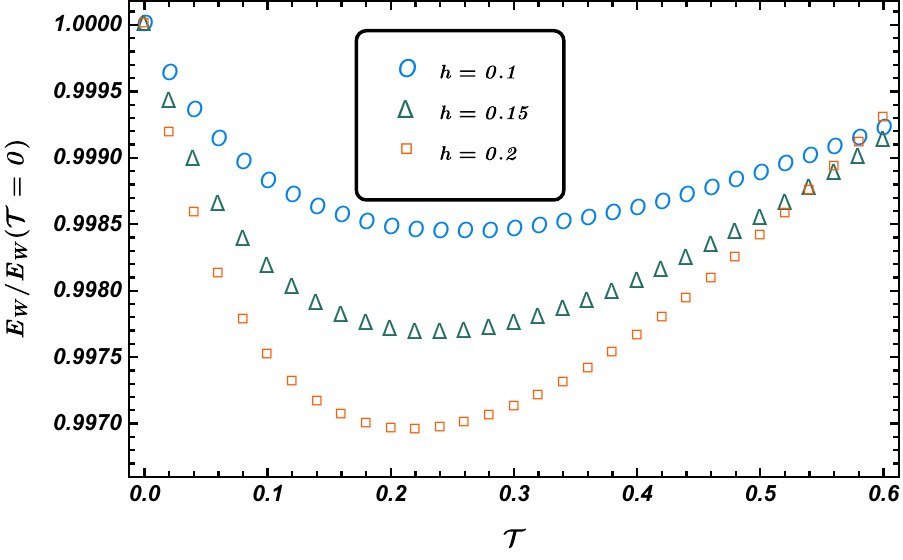}
\end{center}
\caption{EoP as a function of the energy flux for different values of $h$ in $d=3$ with $\ell=1$.}
\label{fig:APW3dcaseA1}
\end{figure}

On the other hand, in $\mathcal{T}\ell^d \ll 1$ limit the metric \eqref{m1} is a small deformation of pure AdS, thus we can use a perturbative expansion to compute the variation of HEE \cite{Mukherjee:2014gia}. Again, in this case $r_t$ is close to the boundary and from eq. \eqref{lAPW} we have
\begin{equation}\label{lrtAPW}
\ell=r_t\left(c+\frac{\sqrt{\pi}}{2(d-1)^2} \left( \frac{\Gamma(\frac{1}{d-1})}{\Gamma(\frac{d+1}{2d-2})}-(d-1) \frac{\Gamma(\frac{d}{2d-2})}{\Gamma(\frac{1}{2d-2})} \right)\mathcal{T}r_t^{d}\right).
\end{equation}
Further, in this limit, the leading order behavior of eq. \eqref{SAPW} reduces to
\begin{equation}\label{dsc}
\Delta S=\frac{R^{d-1}L^{d-2}(d+1)\tilde{c}}{32\pi G_N}\mathcal{T} \ell^2,
\end{equation}
which again shows that HEE increases with $\mathcal{T}$. Combining the above result with eq. \eqref{ds} we can compute HMI in different scaling regimes. In $1\ll h^d \mathcal{T}\ll\ell^d \mathcal{T}$ limit upon substituting eq. \eqref{ds} into eq. \eqref{HMI}, we find
\begin{equation}
I\sim \frac{R^{d-1}L^{d-2}}{G_{N}}  \sqrt{\mathcal{T}}\begin{cases}
\pm  \left(2\ell^{2-\frac{d}{2}}-h^{2-\frac{d}{2}}-(2\ell+h)^{2-\frac{d}{2}}\right)&\qquad d\neq 4\\
\log \left( \frac{\ell^2}{h(2\ell+h)}\right) &\qquad d=4
\end{cases}.
\end{equation}
Subsequently, we can simply use eq. \eqref{dsc} in $h^d \mathcal{T}\ll\ell^d \mathcal{T}\ll 1$ limit which yields 
\begin{align}
\Delta I=-\frac{R^{d-1}L^{d-2}(d+1)\tilde{c}}{16\pi G_N}\mathcal{T} (\ell+h)^2.
\end{align}
This result indicates that HMI decreases when $\mathcal{T}$ is turned on. 
Finally, we turn to the computation of EoP in this background which can be determined by simply evaluating the integral in eq. \eqref{EWCSr} which gives an exact result
\bea\label{ewapw1}
E_W=\frac{R^{d-1}L^{d-2}}{4(d-2)G_N}r^{2-d}{_2}F_1\left(-\frac{1}{2}, \frac{2}{d}-1, \frac{2}{d}, -\frac{\mathcal{T} r^d}{2}\right)\bigg|_{r_u}^{r_d}.
\eea
Now, in $1\ll h^d \mathcal{T}\ll\ell^d \mathcal{T}$ limit, the turning point can be approximated by $r_t\sim \ell$ and we find
\begin{equation}\label{hmiapwcaseA}
E_W\sim \frac{R^{d-1}L^{d-2}}{8G_{N}}  \sqrt{2\mathcal{T}}\begin{cases}
\frac{2}{d-4} \left(h^{2-\frac{d}{2}}-(2\ell+h)^{2-\frac{d}{2}}\right)&\qquad d\neq 4\\
\log \left(1+\frac{2\ell}{h}\right) &\qquad d=4
\end{cases}.
\end{equation}
Interestingly, we see that in this limit the EoP increases with $\mathcal{T}$ which is consistent with the results presented in Figure \ref{fig:APW3dcaseA}. On the other hand in $h^d \mathcal{T}\ll\ell^d \mathcal{T}\ll 1$ limit we use eq. \eqref{lrtAPW} for $r_{u, d}$ to express the EoP as follows
\begin{equation}
\Delta{E_W}=\frac{R^{d-1}L^{d-2}}{G_N}\frac{(d+1)\mathcal{C}}{32\pi} \mathcal{T} \ell(\ell+h),
\end{equation}
where $\mathcal{C}$ is defined in eq. \eqref{mathcalc}. Recall that $\mathcal{C}$ is negative and hence in this limit the EoP decreases as the energy injected to the system increases. Based on this result and eq. \eqref{hmiapwcaseA}, we conclude that the EoP is not a monotonic function of $\mathcal{T}$. This is better shown in figure \ref{fig:APW3dcaseA1} that reports the ratio $E_W/E_W(\mathcal{T}=0)$ versus $\mathcal{T}$ for different values of the parameters in $d=3$. We do not understand what is the reason for this behavior at present and leave this issue for future study.\footnote{Similar results for the behavior of EoP in a different setup has been reported in \cite{Amrahi:2021lgh}.}

\subsection{EWCS in AdS Plane Wave: Case B}\label{APWcaseB}
In this section, we consider extending our analysis to the case where the width of the strip is along the  $x_{d-1}$-direction. Therefore, in terms of the null coordinates we fix the boundary entangling region to be
\begin{equation}
-\frac{\ell}{2\sqrt{2}}\le x^{\pm}\le \frac{\ell}{2\sqrt{2}},\qquad 0\le x_1, \cdots, x_{d-2}\le L.
\end{equation}
The corresponding HEE functional is given by
\begin{equation}
S=\frac{R^{d-1} L^{d-2}}{4G_N} \int dr\frac{\mathcal{L}}{r^{d-1}},\hspace*{2cm}\mathcal{L}=\sqrt{1-2(\partial_{r}x^+) (\partial_{r} x^-)+\mathcal{T} r^d (\partial_r x^+)^2 }. 
\end{equation}
Minimizing the above expression yields the equations of motion for $x^{\pm}(r)$. Since there is no
explicit $x^{\pm}$ dependence, we have two conserved momentums. After some work, the corresponding expression for $\ell$ and $S$ can be simplified as follows \cite{Narayan:2012ks}
\begin{eqnarray}
\ell&=&2\sqrt{2}\int_{0}^{r_t} dr\frac{r^{d-1}}{\sqrt{H(r)}}=-2\sqrt{2}\int_{0}^{r_t}  dr\frac{r^{d-1}(\mathcal{T} r^d -B)}{\sqrt{H(r)}}\label{l1},\\
S&=&\frac{R^{d-1} L^{d-2}}{2G_N} \int_{\epsilon}^{r_t} \frac{\mathrm{d} r}{r^{d-1}} \frac{AB}{\sqrt{H(r)}}\label{sb},
\end{eqnarray}
where $H(r)=\sqrt{A^2 B^2-2Br^{2d-2}+\mathcal{T}r^{3d-2}}$ and $A$ and $B$ are integration constants which can be written as
\bea
A=\frac{r^{d-1}\mathcal{L}}{\mathcal{T} r^d \partial_r x^+ - \partial_r x^-},\hspace*{2cm}B=\frac{\mathcal{T} r^d \partial_r x^+ - \partial_r x^-}{\partial_r x^+}.
\eea
Further, the turning point is determined by
\begin{equation}\label{rs}
\frac{A^2 B^2}{r_t^{2(d-1)}} + \mathcal{T} r_t^d - 2 B=0.
\end{equation}
Next, we turn our attention to evaluating the EoP in this case. The expression takes the same
form as in pure AdS geometry and as a result we obtain
\begin{equation}\label{EWCSb}
E_W=\frac{R^{d-1}L^{d-2}}{4G_{N}}  \int_{r_d}^{r_u} \frac{dr}{r^{d-1}}=\frac{R^{d-1} L^{d-2} }{4(d-2)G_{N}}\left(\frac{1}{r_d^{d-2}}-\frac{1}{r_u^{d-2}}\right).
\end{equation} 
Before examining the behavior of the EoP for general values of the parameters and dimension, it is instructive to analyze the particular case of two dimensional boundary theory, since
the correlation measures can be determined analytically. In fact, this allows us to derive in detail some general features of the EoP for the excited states. We will treat this case separately in the following.

\subsubsection*{A case study: $d=2$}
In this case, using eqs. \eqref{l1} and \eqref{rs} the relation between $r_t$ and $\ell$ as a function of the energy flux can be expressed analytically in closed form as follows \cite{Narayan:2012ks}
\begin{eqnarray}\label{rtlcaseBd2}
\ell\tanh\left(\sqrt{\frac{\mathcal{T}}{2}}\frac{\ell}{2}\right)=\sqrt{2\mathcal{T}}r_t^2.
\end{eqnarray}
For $\mathcal{T}\ell^2\ll 1$, the turning point has the expansion
\bea
r_t=\frac{\ell}{2}\left(1-\frac{\mathcal{T}\ell^2}{48}+\cdots\right),
\eea
which decreases monotonically from its pure AdS value of $\frac{\ell}{2}$ to $0$, as the energy flux is  increased from zero to infinity at fixed $\ell$. Again, this result shows that turning on the perturbation the  extremal hypersurface moves to smaller values of $r$. Now using the above result and eq. \eqref{sb} for HEE, we obtain
\begin{equation}
\Delta S=\frac{R}{4 G_N} \log \frac{\sinh(\sqrt{\frac{\mathcal{T}}{2}}\ell)}{\sqrt{\frac{\mathcal{T}}{2}}\ell}.
\end{equation}
Further, we can expand the variation of HEE for small and large $\mathcal{T}\ell^2$, which yields
\begin{equation}\label{dscaseB}
\Delta S=\frac{R}{4G_N}\begin{cases}
	\sqrt{\frac{\mathcal{T}}{2}}\ell-\log\left(\sqrt{2\mathcal{T}}\ell\right)+\cdots&\qquad \mathcal{T}\ell^2\gg1\\
	\frac{\mathcal{T}\ell^2}{12}-\frac{1}{5}\left(\frac{\mathcal{T}\ell^2}{12}\right)^2+\cdots &\qquad \mathcal{T}\ell^2\ll1
	\end{cases}.
\end{equation}
We note again that the excitations increase the HEE. With these expressions, the variation of HMI can be evaluated as
\begin{equation}
\Delta I=\frac{R}{4G_N}\log\frac{h(2\ell+h)\sinh^2\left(\sqrt{\frac{\mathcal{T}}{2}}\ell\right)}{\ell^2\sinh(\sqrt{\frac{\mathcal{T}}{2}}h)\sinh(\sqrt{\frac{\mathcal{T}}{2}}(2\ell+h))}.
\end{equation} 
As a result, in different scaling regimes we have
\begin{equation}
\Delta I=\frac{R}{4G_N}\begin{cases}
-\frac{\mathcal{T}}{6}(\ell+h)^2+\cdots &\qquad \mathcal{T}h^2\ll \mathcal{T}\ell^2\ll 1\\
	-\log\sqrt{\frac{\mathcal{T}}{2}}\ell-\sqrt{\frac{\mathcal{T}}{2}}h+\cdots &\qquad \mathcal{T}h^2\ll 1\ll \mathcal{T}\ell^2
	\end{cases}.
\end{equation} 
Again, we see that injecting a finite amount of energy into the system decreases the HMI and hence reduce the mutual correlation between subregions. Finally, using eqs. \eqref{EWCSb} and \eqref{rtlcaseBd2}, we obtain the variation of EoP 
\begin{equation}
\Delta E_W=\frac{R}{8G_N}\log\frac{h\tanh\left(\sqrt{\frac{\mathcal{T}}{2}}\frac{2\ell+h}{2}\right)}{(2\ell+h)\tanh\left(\sqrt{\frac{\mathcal{T}}{2}}\frac{h}{2}\right)}.
\end{equation} 
Expanding in different scaling regimes, we thus have
\begin{equation}
\Delta E_W=\frac{R}{8G_N}\begin{cases}
-\frac{\mathcal{T}}{6}\ell(\ell+h)+\cdots &\qquad \mathcal{T}h^2\ll \mathcal{T}\ell^2\ll 1\\
	-\log\sqrt{\frac{\mathcal{T}}{2}}\frac{\ell}{2}-\frac{h}{2\ell}+\cdots &\qquad \mathcal{T}h^2\ll 1\ll \mathcal{T}\ell^2
	\end{cases}.
\end{equation}
Here we see that the EoP is a decreasing function of the energy which is consistent with our
previous results. Further, comparing the above result with eq. \eqref{dscaseB} we see that in the case
of two adjacent intervals and small perturbation we have $\Delta E_W=-\Delta S$ which shows that these measures change with the same rate. This behavior resembles that found in eq. \eqref{dsdewd2} where the boundary excited state is an equilibrium state at finite temperature.



\subsubsection*{Correlation Measures in Higher Dimensions}
\begin{figure}[h]
\begin{center}
\includegraphics[scale=0.59]{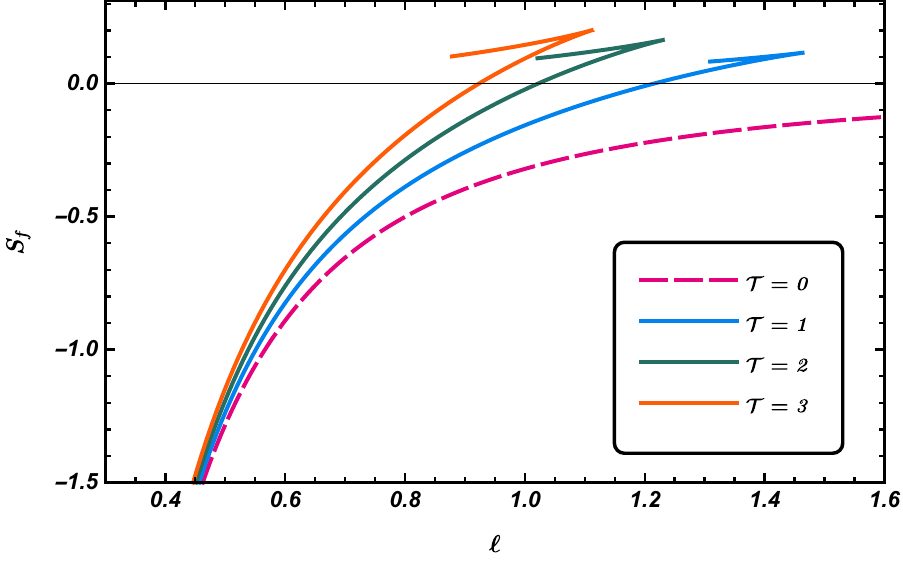}
\hspace*{0.04cm}
\includegraphics[scale=0.58]{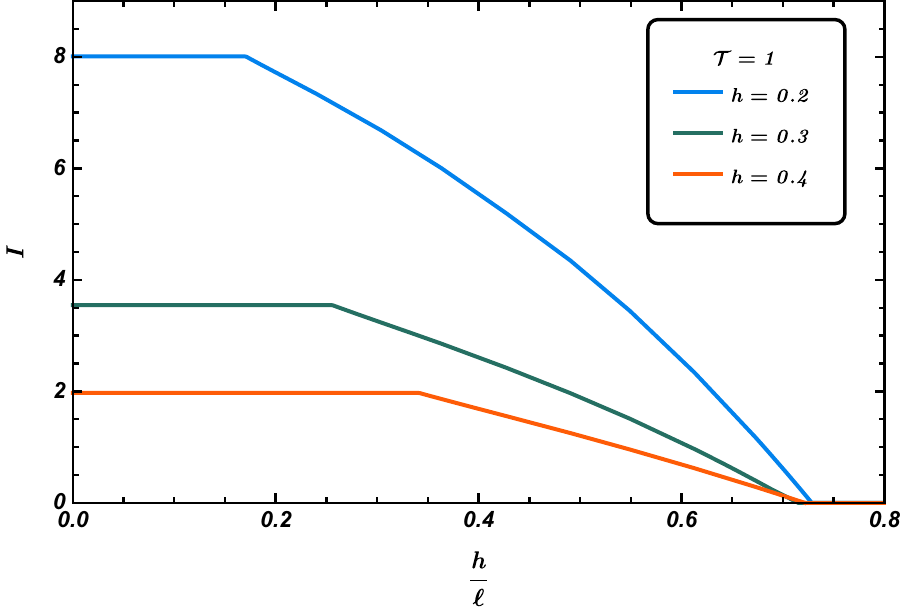}
\hspace*{0.04cm}
\includegraphics[scale=0.58]{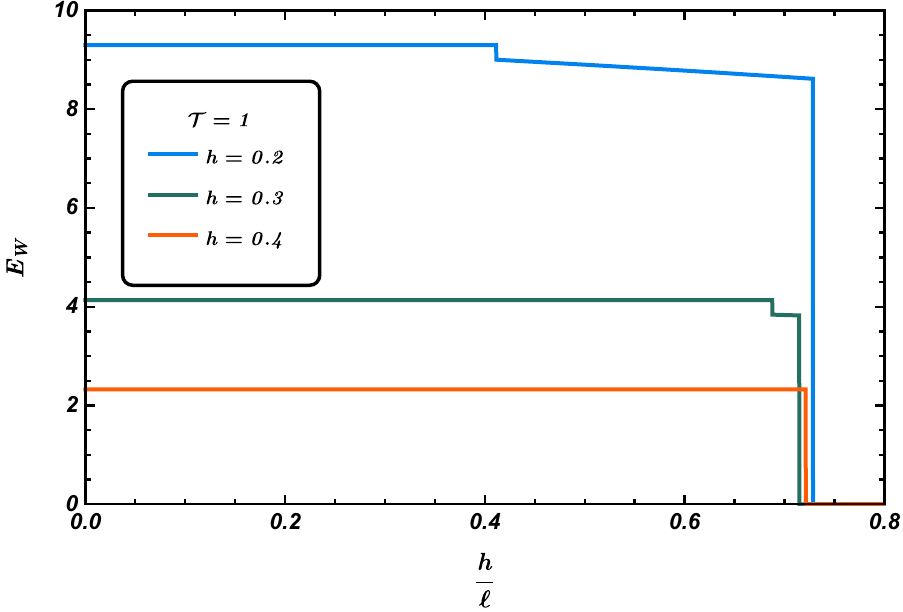}
\end{center}
\caption{Finite part of the HEE (left), HMI (middle) and EWCS (right) as functions of the width and separation between subregions for different values of $\mathcal{T}$ and $h$ in $d=4$.}
\label{fig:APW4d}
\end{figure}

The integrals in eqs. \eqref{l1} and \eqref{sb} cannot be carried out analytically for general $d > 2$, so we turn our attention to numerically evaluating the correlation measures in $d=4$ noting that the qualitative features of the measures are independent of the dimensionality of the field theory. Indeed, as shown in \cite{Narayan:2012ks} in $d\geq 3$ there are two candidates for the minimal hypersurface with the same endpoints at the asymptotic boundary. The first one is the union of two disconnected straight lines extending from
$r=\epsilon$ to $r=\infty$. In this case we have $\partial_r x^{\pm}=0$ and the corresponding HEE is independent of $\ell$. The second is a curved surface which extends from $r=\epsilon$ to $r=r_t$ and whose area depends on $\ell$. In general, there is a critical length $\ell_c$ such that for $\ell<\ell_c$ the connected configuration has the minimal area, while for $\ell>\ell_c$ the RT hypersurface changes topology and the disconnected configuration is favored. Note that in this case the finite part of the HEE is given by $S_f=S_{\rm con.}-S_{\rm dis.}$. The behavior of correlation measures can be read off from Fig. \ref{fig:APW4d}. The left panel shows the finite part of the HEE as a function of the width of the entangling region for different values of $\mathcal{T}$. From this figure one finds that the critical width when the transition happens is approximately given by $\mathcal{T}^{1/4}\ell_c\sim 1.17$. Note that the dashed curve corresponds to pure AdS geometry with $\mathcal{T}=0$ where the connected configuration is always favored for any $\ell$. The middle panel demonstrates the HMI as a function of $\frac{h}{\ell}$ for $\mathcal{T}=1$. In the right panel we show $E_W$ for the same values of the parameters.

We can also examine the behavior of the EoP in $\mathcal{T}\ell^d\ll 1$ limit. The analysis follows identically to the previous cases, hence we only present the results. In this limit we can expand eqs. \eqref{l1} and \eqref{sb} to find
\begin{equation*}
\Delta S=\frac{R^{d-1} L^{d-2} \tilde{c}}{16\pi G_N}\mathcal{T} \ell^2,
\end{equation*}
where $\tilde{c}$ is defined in eq. \eqref{tildec}. We see that the variation of HEE is positive and monotonically increasing as a function of $\mathcal{T}$. Using this result one can show that the corresponding variation of HMI is negative as expected. Also expanding eq. \eqref{EWCSb} to leading order yields
\begin{equation*}
\Delta E_W=\frac{R^{d-1}L^{d-2} \mathcal{C}}{16\pi G_N} \mathcal{T} \ell(\ell+h),
\end{equation*}
where $\mathcal{C}$ is defined in eq. \eqref{mathcalc}. Again, in this limit the EoP decreases as the energy injected to the system increases. Note that similar to the previous case our numerical results show that the EoP is not a monotonic function of the energy flux, although we do not explicitly show the corresponding figures here.

 \section{Momentum Relaxation Geometries}\label{MomentumRelax}
%
%
%
%
%
%
%
%
%
In this section, we consider a specific class of boundary excited states in which a scalar operator acquiring an expectation value. Based on holographic dictionary the dual description involves adding a scalar field to the bulk. In particular, we focus on a model of translational symmetry breaking in which inhomogeneous
scalar profiles can be constructed in such a way that the corresponding geometry remains isotropic and homogeneous. In this model the gravitational theory consists of Einstein-Maxwell theory with minimally coupled massless scalar fields. As shown in \cite{Andrade:2013gsa}, the spatially-dependent marginal scalar operators cause momentum relaxation in the deformed dual field theory. The corresponding metric in $d>2$ reads
\begin{eqnarray}\label{relaxmetric}
ds^2=\frac{R^2}{r^2}\left(-f(r)dt^2+\sum_{i=1}^{d-1}dx_i^2+\frac{dr^2}{f(r)}\right),\hspace*{1cm}f(r)=1-mr^d-\frac{\alpha^2 r^2}{2(d-2)},
\end{eqnarray}
where $\alpha$ is a constant which is proportional to the strength of the source for the boundary scalar operator. From \eqref{relaxmetric}, one obtains that the Hawking temperature and the horizon radius can be specified by
\begin{eqnarray}
T=\frac{d}{4\pi r_h}\left(1-\frac{\alpha^2 r_h^2}{2d}\right),\hspace*{2cm}m=\frac{1}{r_h^d}\left(1-\frac{\alpha^2 r_h^2}{2(d-2)}\right).
\end{eqnarray}
On the other hand for $d=2$ the blackening factor becomes logarithmic, \textit{i.e.}, 
\begin{eqnarray}
f(r)=1-mr^2+\frac{\alpha^2 r^2}{2}\log r.
\end{eqnarray}
Further, the extremal limit is obtained with $\alpha^2r_h^2=2d$ and
also by setting
\bea\label{extmetric}
m=\Bigg\{ \begin{array}{rcl}
\frac{2}{(2-d)r_h^d}&{\rm for}&d>2\\
\frac{1+2\log r_h}{r_h^2}&{\rm for}&d=2
\end{array}.
\eea
Before we proceed, let us recall that different aspects of non-local probes including HEE and HMI in this model has been studied in \cite{Mozaffara:2016iwm}. Indeed, as shown in this paper for $d>2$ due to momentum relaxation effects, new logarithmic universal terms may appear in the HEE. On the other hand, in the case of a two dimensional boundary theory, the momentum relaxation leads to a non-critical correction and the universal term remains unchanged. Further, the correlation length is a decreasing function of momentum relaxation parameter and hence the phase transition of HMI happens at smaller separation between the spatial subregions.

Regarding the above observations we would like to study the effect of momentum relaxation on EoP. As before, we consider a symmetric configuration consisting of two parallel strips. The analysis follows identically to the previous cases discussed in section \ref{adsbb}, with the replacement of the blackening factor. Therefore, the corresponding expressions for the width of the entangling region and HEE are given by eqs. \eqref{Xadsbb} and \eqref{heestat} respectively. 
Further the EoP functional still takes the form presented in eq. \eqref{ewbb}. We copy it here for convenience,
\begin{align}\label{ewMR}
E_W=\frac{R^{d-1}L^{d-2}}{4G_N}\int_{r_d}^{r_u} \frac{dr}{r^{d-1}\sqrt{f(r)}}.
\end{align}
Before examining the full $\alpha$-dependence of EoP, we would like to study the variation of this quantity in more detail in $m=0, h\ll \ell\ll \alpha^{-1}$ regime which corresponds to the small deformation of the boundary state with scalar condensation at finite temperature. Of course, the overall conclusions are independent of the specific value of $m$, but we will focus on this regime for concreteness. In this case, evaluating the above integral yields
\begin{align}\label{ewMR1}
E_W=\frac{R^{d-1}L^{d-2}}{4(2-d)G_N}\frac{1}{r^{d-2}} \,_{2}F_1\left(\frac{1}{2}, 1-\frac{d}{2}, 2-\frac{d}{2}, \frac{\alpha^2 r^2}{2(d-2)}\right)\bigg|_{r_d}^{r_u}.
\end{align}
Now, expanding the above result in $h\ll \ell\ll \alpha^{-1}$ limit which corresponds to $r_d\ll r_u\ll \alpha^{-1}$ we have
\begin{align}
E_W=\frac{R^{d-1}L^{d-2}}{4(d-2)G_N}\left(\frac{1}{r_d^{d-2}}-\frac{1}{r_u^{d-2}}+\frac{\alpha^2}{4(d-4)}\left(\frac{1}{r_d^{d-4}}-\frac{1}{r_u^{d-4}}\right)\right).
\end{align}
It is straightforward to write the above expression in terms of the boundary quantities $h$ and $\ell$ to produce the result
\begin{align}\label{ewMRT0}
\Delta E_W=\frac{R^{d-1}L^{d-2}c^{d-4}(1+\mathcal{C}')}{16(d-2)(d-4)G_N}\left(\frac{1}{h^{d-4}}-\frac{1}{(2\ell+h)^{d-4}}\right)\alpha^2,
\end{align}
with
\begin{eqnarray}\label{mathcalcprim}
\mathcal{C}'=\frac{2\sqrt{\pi}(d-4)}{(d+2)c}\frac{\Gamma\left(\frac{3d}{2d-2}\right)}{\Gamma\left(\frac{2d+1}{2d-2}\right)}.
\end{eqnarray}
Note that the overall coefficient in eq. \eqref{ewMRT0} is positive for $d>4$ and hence the scalar condensation can increase the EoP. We also note that this result does not hold in $d=2, 4$. It is easy to show that in $h\ll \ell\ll \alpha^{-1}$ limit we have
\bea
\Delta E_W=\frac{R^{d-1}L^{d-2}}{32G_N}\alpha^2\Bigg\{ \begin{array}{rcl}
\frac{1}{2}\log \frac{\ell}{h}&{\rm for}&d=4\\
\frac{1}{3}\log \alpha \ell&{\rm for}&d=2
\end{array}.
\eea
Based on the above expression and eq. \eqref{ewMRT0}, we conclude that the variation of EoP is positive for $d>3$. It is worth to mention that considering the extremal limit which corresponds to a boundary state at zero temperature, the qualitative features of the above results do not change.

To close this section, we present plots of HEE, HMI and EoP in fig. \ref{fig:BBBcaseB} for several values of the momentum relaxation parameter in a three dimensional boundary theory. 
In the left panel, we present the HEE as a function of the width of the entangling region. Clearly it is an increasing function of the momentum relaxation parameter in accordance with the previous semi-analytic results \cite{Mozaffara:2016iwm}. The middle panel demonstrates HMI as a function of $h/\ell$ which undergoes a
continuous phase transition beyond which it is identically zero. In the right panel we show EoP for the same
values of the parameters which has a discontinuous phase transition. Our numerical results make it clear that in this case the phase transition of HMI and EoP happens at smaller separation between the spatial subregions compared to pure AdS case with $\alpha=0$. 
\begin{figure}
\begin{center}
\includegraphics[scale=0.595]{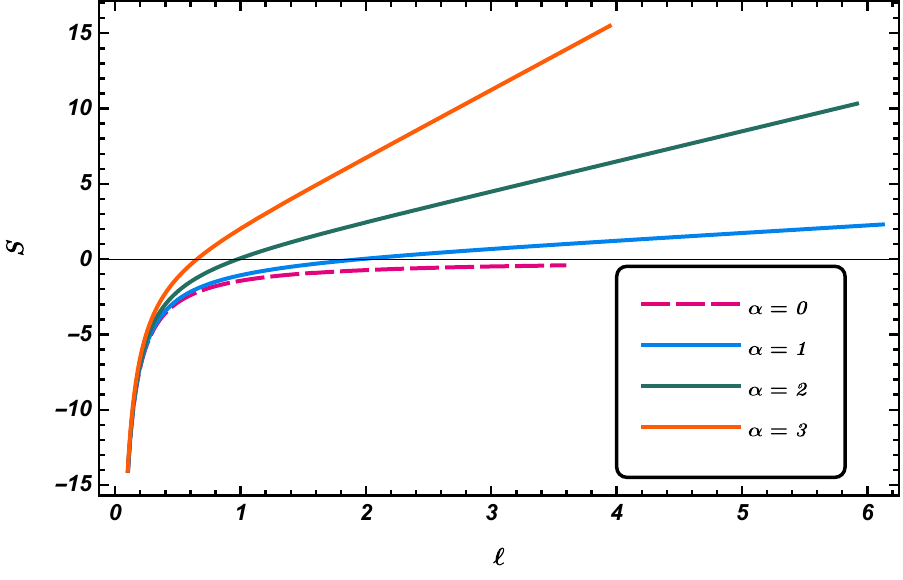}
\hspace*{0.04cm}
\includegraphics[scale=0.575]{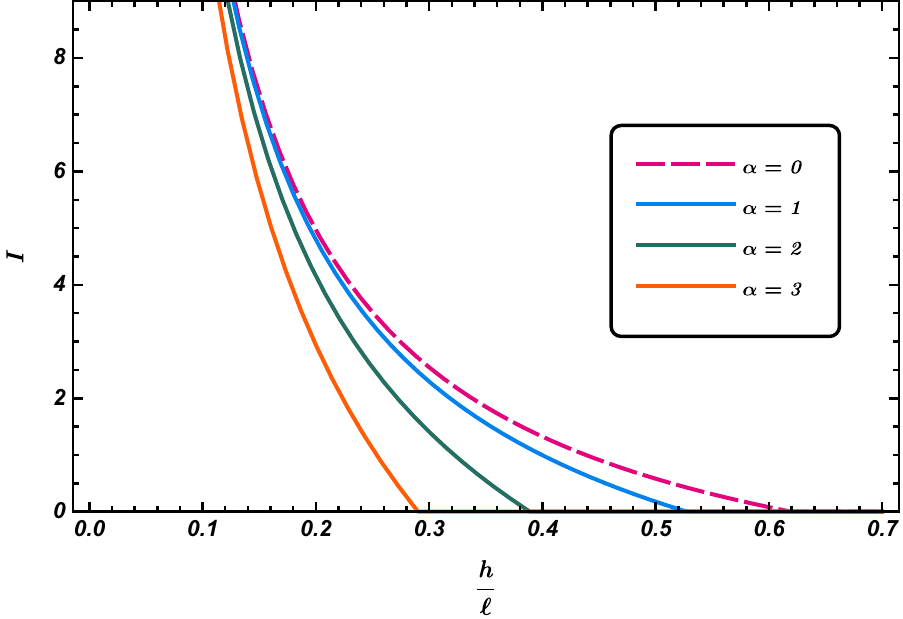}
\hspace*{0.04cm}
\includegraphics[scale=0.575]{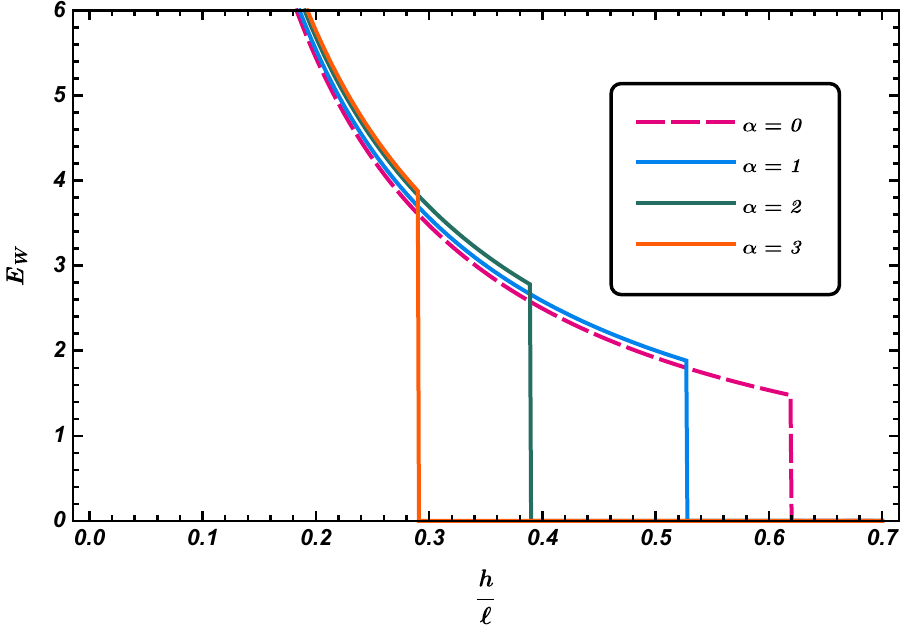}
\end{center}
\caption{The HEE (left), HMI (middle) and EoP (right) as a function of $h$ and $\ell$ for different
values of the momentum relaxation parameter. Here we have set $d=3$ and $m=0$.}
\label{fig:MR}
\end{figure}

\section{General Analysis}\label{general}

In this section we extend our studies to general boundary excited states
which are dual to asymptotically AdS geometries. We start with the asymptotically AdS$_{d+1}$ metric in Fefferman-Graham (FG) coordinates\footnote{Note that in the following, $z$ denotes the bulk radial coordinate in a generic asymptotically AdS geometry, hence for the pure AdS spacetime $z=r$ as in previous sections.}
\begin{eqnarray}\label{FG}
ds^2=\frac{R^2}{z^2}\left(dz^2+g_{\mu\nu}(z, x^\rho)dx^\mu dx^\nu\right)\hspace*{2cm}g_{\mu\nu}(z, x^\rho)=\eta_{\mu\nu}+\gamma_{\mu\nu}(z, x^\rho),
\end{eqnarray}
where for small deviations from the vacuum we consider $|\gamma_{\mu\nu}|\ll 1$. 
In the following, we are interested in the first order correction of EoP under a small perturbation. 
While this analysis can be done for a generic entangling surface, in order to gain better insight into the properties of EoP, let us illustrate the discussion with the example of a configuration consisting of two parallel strips with arbitrary widths along, say, the direction of $x_1\equiv x$. As in the previous sections, the symmetry of our setup implies that the corresponding profile for the minimal hypersurface $\Sigma$ parametrized as $x=x(z)$. Note that, for pure AdS geometry, \textit{i.e.}, $\gamma_{\mu\nu}=0$, the corresponding hypersurface lies entirely on a constant time slice. Hence, we assume that in the perturbed geometry, the profile for the minimal hypersurface can be given as follows
\begin{eqnarray}\label{profilepert}
x(z)=x_0(z)+x_1(z)+\cdots.
\end{eqnarray}
In this case using eq. \eqref{FG}, the EoP functional can be written as
\begin{eqnarray}\label{ewpert}
E_W=\frac{R^{d-1}L^{d-2}}{4G_N}\int_{\bar{z}_d}^{\bar{z}_u} \frac{dz}{z^{d-1}}\sqrt{g}\sqrt{g^{xx}+x'^2},
\end{eqnarray}
where $g\equiv \det (g_{ij})$. Here $\bar{z}_{u, d}$ denote the values of $z$ at the intersection of $\Sigma$ and $\Gamma$s in the perturbed background which coincide with $z_{u, d}$ for $\gamma_{\mu\nu}=0$. Thus to leading order
\begin{eqnarray}
\bar{z}_{u, d}={z}_{u, d}+{z}^{(1)}_{u, d}+\cdots.
\end{eqnarray}
Using eq. \eqref{FG} the first factor inside the integral in eq. \eqref{ewpert} becomes
\begin{eqnarray}
\sqrt{g}=\sqrt{1+\gamma_i^i}=1+\frac{1}{2}\gamma_i^i,
\end{eqnarray}
where $\gamma_i^i=\delta_{ij}\gamma^{ij}$. We can also expand the final factor in eq. \eqref{ewpert} as follows
\begin{eqnarray}
\sqrt{g^{xx}+x'^2}=\sqrt{1+x_0'^2}\left(1-\frac{\gamma_{xx}-2x_0'x_1'}{2(1+x_0'^2)}\right).
\end{eqnarray}
Combining the above results, we obtain the  variation of EoP
\begin{eqnarray}\label{eopvar}
\Delta E_W=\frac{R^{d-1}L^{d-2}}{4G_N}\left(\frac{z_u^{(1)}\sqrt{1+x_0'(z_u)^2}}{z_u^{d-1}}-\frac{z_d^{(1)}\sqrt{1+x_0'(z_d)^2}}{z_d^{d-1}}+\int_{z_d}^{z_u} \frac{dz\sqrt{1+x_0'^2}}{2z^{d-1}}\left(\gamma^{i}_i-\frac{\gamma_{xx}-2x_0'x_1'}{1+x_0'^2}\right)\right),
\end{eqnarray}
where we subtracted the vacuum contribution which is in this case given by
\begin{eqnarray}
E_{W {\rm AdS}}=\frac{R^{d-1}L^{d-2}}{4G_N}\int_{z_d}^{z_u} \frac{dz}{z^{d-1}}\sqrt{1+x_0'^2}.
\end{eqnarray}
Turning now to the symmetric configuration where the width of the subregions are equal and $\Sigma$ lies entirely on $x_0=0$ slice, \textit{i.e.}, $x_0'=0$, eq. \eqref{eopvar} can be rewritten as follows
\begin{eqnarray}\label{eopvar1}
\Delta E_W=\frac{R^{d-1}L^{d-2}}{4G_N}\left(\frac{z_u^{(1)}}{z_u^{d-1}}-\frac{z_d^{(1)}}{z_d^{d-1}}\right)+\frac{R^{d-1}L^{d-2}}{8G_N}\int_{z_d}^{z_u} \frac{dz}{z^{d-1}}\left(\gamma^{i}_i-\gamma_{xx}\right).
\end{eqnarray}
Notice that the first term in the above expression comes from the variation of the RT hypersurfaces, \textit{i.e.}, the boundary condition for EWCS, while the second term is the contribution coming from the variation of EoP functional. 

The above calculations hold for a generic  perturbation
of the vacuum state at leading order. Recall that the perturbation of the boundary vacuum state described by different excitations in the dual geometry. Indeed, the deviation of the bulk metric from pure AdS is given by \cite{Blanco:2013joa}
\begin{eqnarray}\label{gamma}
\gamma_{\mu\nu}=\frac{16\pi G_N}{dR^{d-1}}z^d\sum_{n=0}z^{2n}\;T_{\mu\nu}^{(n)},
\end{eqnarray}
where using the equations of motion one can show that $T^{(n)}_{\mu\nu}$ is
traceless and conserved, \textit{i.e.}, 
\begin{eqnarray}
T^{(n)\mu}_{\mu}=0,\hspace*{2cm}\partial_\mu T^{(n)\mu}_{\nu}=0.
\end{eqnarray}
Further, employing the Einstein equations, $T_{\mu\nu}^{(n>0)}$ can be determined in terms of $T_{\mu\nu}^{(0)}$, \textit{i.e.}, $T^{(n)}_{\mu\nu}\propto \Box^n T^{(0)}_{\mu\nu}$. Throughout the following, we will assume that the boundary excitations are uniform and hence $T^{(n>0)\mu\nu}=0$. Equipped with eq. \eqref{eopvar1}, we evaluate the variation of EoP for three different kinds of excitations in more detail in the following. Let us recall that similar analysis for the variation of HEE when the entangling subregion is a sphere has been done in \cite{Blanco:2013joa}.

\subsection*{Case i) $\langle T_{\mu\nu}\rangle \neq 0$}\label{gravityper}
Let us first focus our attention on the boundary states described by purely gravitational excitations in the dual geometry. In this case the stress tensor is the only operator that has a nonvanishing expectation value. 
%
Assuming that $\langle T_{\mu\nu}\rangle\sim\delta\ll 1$, the change in the EoP may receive contributions at all orders in $\delta$.\footnote{Note that we introduce $\delta \ll 1$ to control the overall amplitude of the boundary excitations which allows us to easily keep track of the perturbative expansion to all orders. Hence after evaluating the corrections we set $\delta=1$.} However, in the following, our
calculations will be to linear order in $\delta$. 

Notice that, we would like to find the variation of the EoP in terms of boundary quantities $\ell$ and $h$. We
do so by first finding the leading correction to the turning point of the RT hypersurface. It is relatively straightforward to evaluate this correction to linear order in $\delta$, with the result
\begin{eqnarray}\label{varewzt}
z_t=\frac{\ell}{c}+\frac{8\pi G_N}{d(d-1)R^{d-1}}\left(T_{00}-\left(T_{00}-\frac{d-1}{d+1}T_{xx}\right)\frac{2\sqrt{\pi}}{c}\frac{\Gamma\left(\frac{d}{d-1}\right)}{\Gamma\left(\frac{d+1}{2d-2}\right)}\right)\left(\frac{\ell}{c}\right)^{d+1},
\end{eqnarray}
where we have used $T^{i}_i=T_{00}$ due to the tracelessness of the stress tensor. Upon substituting eqs. \eqref{gamma} and \eqref{varewzt} into eq. \eqref{eopvar1} we may evaluate the variation of EoP to find
\begin{eqnarray}\label{eopvar12}
\Delta E_W\left(T_{\mu\nu}\right)=\frac{\mathcal{C}}{d}L^{d-2}\bigg((d+1)T_{00}-(d-1)T_{xx}\bigg)\ell(\ell+h),
\end{eqnarray}
where $\mathcal{C}$ is defined in eq. \eqref{mathcalc}. Clearly, the first term above is completely determined by the energy density. Whereas the second term receives contribution from the $xx$ component of the stress tensor. Precisely, the same situation arose in \cite{Blanco:2013joa} in investigating the structure of the first order variation of the HEE. There it was shown that although for spherical subregions only the energy density contributes to the variation of HEE, for entangling surfaces with a less symmetric geometry, other components have nontrivial contributions. For example, if we consider a strip entangling region, the variation of HEE becomes \cite{Blanco:2013joa}
\begin{eqnarray}\label{heevargeneral}
\Delta S\left(T_{\mu\nu}\right)=\frac{\tilde{c}}{d}L^{d-2}\bigg((d+1)T_{00}-(d-1)T_{xx}\bigg)\ell^2,
\end{eqnarray}
where $\tilde{c}$ is defined in eq. \eqref{tildec}. For the case of $d=2$, comparing the above result to eq. \eqref{eopvar12} shows that for adjacent intervals at leading order $\Delta E_W=-\Delta S$. Further, in the case of an isotropic background, \textit{i.e.}, $T_{xx}=T^{i}_i/(d-1)$, eq. \eqref{eopvar12} yields 
\begin{eqnarray}\label{eopvar2}
\Delta E_W=\mathcal{C}L^{d-2}\ell(\ell+h)T_{00}.
\end{eqnarray}
Recall that $\mathcal{C}<0$ and hence the above result demonstrates that the excitations decrease the EoP. One simple consistency check on our result is that in the case of AdS black brane geometry, we recover the expected results eqs. \eqref{ewvarBB} and \eqref{ewbbext}. We would also like to stress that for the present case, the variation of HEE is always positive. More explicitly, using eq. \eqref{heevargeneral} the leading correction to HEE for isotropic backgrounds is given by
\begin{eqnarray}\label{heestress}
\Delta S=\tilde{c}L^{d-2}\ell^2 T_{00}.
\end{eqnarray}
This result indicates that the corresponding HMI decreases, similar to what was observed above for the EoP.
\subsection*{Case ii) $\langle T_{\mu\nu}\rangle \neq 0,\;\langle J_{\mu}\rangle \neq 0$}\label{currentper}
Suppose the boundary theory has a conserved $U(1)$ current $J^\mu$ such that $\langle J^{0}\rangle$ is the corresponding charge density. This means that there should be a massless gauge field $A_\mu$ in the bulk whose boundary value determines the chemical potential conjugate to $\langle J^{0}\rangle$. Taking into account the leading backreaction of the gauge field on the geometry, the relevant part of the metric perturbation in the FG expansion will take the form \cite{Blanco:2013joa}
\begin{eqnarray}
\gamma_{\mu\nu}=\frac{16\pi G_N}{dR^{d-1}}z^d T_{\mu\nu}+\frac{z^{2(d-1)}}{4(d-1)^2(d-2)}\left(2(1-d)J_\mu J_\nu+\eta_{\mu\nu}J^2\right).
\end{eqnarray}
Note that the leading order contribution to EoP coming from the stress tensor is given by eq. \eqref{eopvar12} and so in the following, we focus on the contribution due to the second term above. Assuming an isotropic current density, \textit{i.e.}, $J_x^2={\vec{J}}^{\,2}/(d-1)$, we may evaluate the variation in eq. \eqref{eopvar1} to find
\begin{eqnarray}\label{eopvarcurrent}
\Delta E_W\left(J^{\mu}\right)=\frac{R^{d-1}L^{d-2}}{16G_Nc^d}\frac{d-1}{d(d-2)(2d-1)}\left((J^0)^2+{\vec{J}}^{\,2}\right)\left((2\ell+h)^d-h^d\right).
\end{eqnarray}
This result indicates that EoP increases when the current density is turned on. Hence we find that the correlation between subsystems is enhanced by adding the chemical potential. On the other hand one can evaluate the variation of HEE due to the current as follows
\begin{eqnarray}\label{heevarcurrent}
\Delta S\left(J^{\mu}\right)=-\frac{R^{d-1}L^{d-2}}{16G_Nc^{d-1}}\frac{1}{(d-2)(2d-1)}\left((J^0)^2+{\vec{J}}^{\,2}\right)\ell^d,
\end{eqnarray}
which is always negative. Based on this result one can show that the current perturbations
 produce a positive contribution to the change in the HMI which is consistent with the behavior of EoP. Finally, note that the qualitative features of the above results are in complete agreement with eqs. \eqref{deltasq} and \eqref{ewbbext}.

\subsection*{Case iii) $\langle T_{\mu\nu}\rangle \neq 0,\;\langle\mathcal{O}\rangle \neq 0$}\label{scalarper}
\begin{figure}
\begin{center}
\includegraphics[scale=0.75]{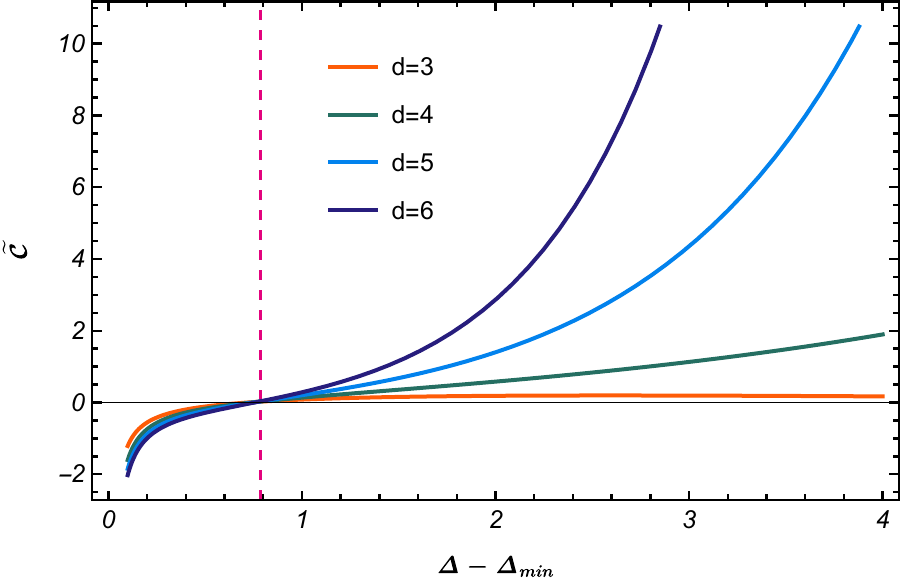}
\end{center}
\caption{$\widetilde{\mathcal{C}}$ as a function of scaling dimension for different
values of $d$. The vertical dashed line at $\Delta-\Delta_{\rm min}\sim \frac{\pi}{4}$ indicates the critical point where $\Delta E_W\left(\mathcal{O}\right)$ flips its sign.}
\label{fig:widetildeC}
\end{figure}

Now consider perturbing the boundary theory by adding a scalar operator with scaling dimension $\Delta$. This boundary operator is dual to a massive bulk scalar field whose mass should satisfy $m=\Delta(d-\Delta)$. Note that the unitarity implies a lower bound on $\Delta$, \textit{i.}{e.}, $\Delta>\Delta_{\rm min}=\frac{d}{2}-1$. In the absence of any sources the leading backreaction of the scalar field on the geometry in the FG expansion will take the form \cite{Blanco:2013joa}
\begin{eqnarray}
\gamma_{\mu\nu}=\frac{16\pi G_N}{dR^{d-1}}z^d T_{\mu\nu}-\frac{\mathcal{N}^2}{4(d-1)}z^{2\Delta}\eta_{\mu\nu} \mathcal{O}^2,
\end{eqnarray}
where $\mathcal{N}$ is a normalization constant and $\mathcal{O}$ denotes the scalar condensate. Again, the leading order contribution coming from the stress tensor is given by eq. \eqref{eopvar12} and so in the following, we focus on the contribution due to the second term. It is relatively straightforward to evaluate the leading correction to EoP for a uniform scalar condensate with the result
\begin{eqnarray}\label{scalardew}
\Delta E_W\left(\mathcal{O}\right)=\frac{R^{d-1}L^{d-2}\widetilde{\mathcal{C}}}{16G_N}\left((2\ell+h)^{2\Delta-d+2}-h^{2\Delta-d+2}\right)\mathcal{N}^2 \mathcal{O}^2,
\end{eqnarray}
where
\begin{eqnarray}
\widetilde{\mathcal{C}}=\frac{-\Delta}{(d-1)(2\Delta-d+2)c^{2\Delta-d+2}}\left(1-\sqrt{\pi}\frac{2\Delta-d+2}{(d-1)c}\frac{\Gamma\left(\frac{2\Delta+d}{2d-2}\right)}{\Gamma\left(\frac{2\Delta+2d-1}{2d-2}\right)}\right).
\end{eqnarray}
Using the unitarity bound, one can show that $\widetilde{\mathcal{C}}$ flips its sign at a critical point which can be approximated by $\Delta-\Delta_{\rm min}\sim \frac{\pi}{4}$ (see figure \ref{fig:widetildeC}). Thus for a sufficiently relevant scalar operator $\Delta E_W\left(\mathcal{O}\right)$ is negative which means that the scalar condensate decreases the EoP. Note that the total variation of the EoP for an excited state with nontrivial expectation values for stress tensor and scalar operator can be obtained by combining eqs. \eqref{eopvar12} and \eqref{scalardew}. Assuming that the magnitude of $T_{\mu\nu}$ and $\mathcal{O}$ is controlled by a single scale $M$ we have
\begin{eqnarray}\label{compare}
\frac{\Delta E_W\left(\mathcal{O}\right)}{\Delta E_W\left(T_{\mu\nu}\right)}\sim \left(M\mathcal{L}\right)^{2\Delta-d}.
\end{eqnarray}
where $\mathcal{L}$ is a length scale depending on $h$ and $\ell$. Since we are working in a regime where
$M\mathcal{L}\ll 1$, for $\Delta<\frac{d}{2}$ the contribution due to the scalar condensate is dominant.
Further, one can compute the variation of HEE in the present case as follows 
\begin{eqnarray}\label{scalards}
\Delta S\left(\mathcal{O}\right)=-\frac{R^{d-1}L^{d-2}}{16G_N}\frac{\Delta}{(d-1)(2\Delta-d+2)c^{2\Delta-d+2}}\frac{\sqrt{\pi}\Gamma\left(\frac{2\Delta+d}{2d-2}\right)}{\Gamma\left(\frac{2\Delta+2d-1}{2d-2}\right)}\ell^{2\Delta-d+2}\mathcal{N}^2 \mathcal{O}^2.
\end{eqnarray}
Again, considering the unitarity bound, the overall factor in the above expression is negative and hence turning on the scalar operator decreases the HEE. Also note that based on the above result the variation of HMI is positive. This behavior contrast with the corresponding result for the variation of EoP when $\mathcal{O}$ is sufficiently relevant. Interestingly, while both HMI and EoP are measures of total correlation between subregions, they do not behave in the same manner as the vacuum state perturbed by a scalar operator. We do not fully understand what is the reason for this behavior and leave it for future study. Note that based on figure \ref{fig:widetildeC} for sufficiently irrelevant operators, the EoP and HMI behave in the same manner.

\section{Conclusions and Discussions}\label{diss}

In this paper, we explored the behavior of entanglement wedge cross section (EWCS) in different asymptotically AdS geometries dual to boundary excited states. We used the holographic proposal established in \cite{Takayanagi:2017knl} for computing this quantity which gives $E_W$ in terms of the minimal cross-sectional area of the entanglement wedge. In particular, we computed the variation of EWCS under the small perturbations away from the vacuum, when generic operators acquire nontrivial expectation values. To get a better understanding of the results, we also compared the variation of $E_W$ to other correlation measures including HEE and HMI. Our study was mainly for a symmetric configuration consisting of two disjoint strips with equal width, which is the simplest case to utilize the holographic proposal to compute the mixed state correlation measures. However, we expect that the qualitative features of our results are independent of the specific configuration. Although, for finite excitations we did a numerical analysis, considering small perturbations around the vacuum, we evaluated the leading order variation of holographic correlation measures analytically. 

A key observation was that the variation of RT hypersurfaces plays a central role in evaluating the correction to EWCS even at leading order. This is completely different from what happens for the HEE (or HMI) in the same setup, where any change in the profile of the RT hypersurface does not contribute to the first order correction.

Our analysis in this paper focused mainly on boundary excited states described by
purely gravitational excitations in the bulk where the stress tensor is the only operator that has a
nonvanishing expectation value. Generally, the leading correction to $E_W$ is negative which means that the excitations decrease the correlation between the subregions and hence promote disentangling between them. 
Of course, this is in agreement with the previous results for the variation of HMI in the same setup \cite{Alishahiha:2014jxa}. Regarding the EWCS and HMI as measures of total correlation between the subregions, this behavior seems reasonable. However, this result is different from what happens for the HEE  which an increasing function of the excitation parameter due to the fact that the number of degrees
of freedom grows as we excite the system. These results hold in different excited states dual to perturbed bulk geometries including AdS black brane and AdS plane wave backgrounds. Further, correction to EWCS in a general asymptotically AdS geometry with purely gravitational excitations is given in eq. \eqref{eopvar12}. In this case, we found that based on the symmetry of the entangling regions a combination of the components of the stress tensor contribute to the variation of $E_W$. 

In extending these calculations to states in which additional matter fields are excited in the dual geometry, we considered two different types of excitations: perturbing with a current and perturbing with a scalar condensate. Upon adding the extra matter fields, a number of interesting features arose. First, the EWCS increases when the current density is turned on and hence the total correlation between the subregions increased. This is consistent with the variation of HMI being positive in this case. Second, the scalar condensate produces a negative contribution to the change in the EWCS, when the scalar operator is sufficiently relevant and its dimension respects the unitarity bound. Further, comparing the contributions to $\Delta E_W$ due to the stress tensor and scalar operator, we see that when the scaling dimension of the scalar operator satisfy $\Delta<d/2$, the contribution due to the scalar condensate is dominant.

At this point, let us recall that there exist different correlation measures which can be regarded as the holographic dual to EWCS, \textit{e.g.}, entanglement of purification (EoP), reflected entropy and odd entropy. Given the holographic prescriptions in eqs. \eqref{ewep} and \eqref{ewsr}, it is clear that all the above results, up to overall multiplicative constant, hold for the EoP and reflected entropy. However, the situation is different for odd entropy where the corresponding expression is given by eq. \eqref{odd}. Remember that we always consider the regime where the connected configuration is favored, \textit{i.e.}, $S_{A\cup B}=S(2\ell+h)+S(h)$, it is straightforward to compute the odd entropy using our previous results for different excited states. In general, the explicit expression for $S_O$ is somewhat complicated, so we will not write it out here. However, when $A$ and $B$ are two adjacent intervals in an excited state of a 2-dimensional CFT, the resultant odd entropy has a rather simple expression. In particular, when $T_{\mu\nu}$ is the only operator that has a nonvanishing
expectation value, using eqs. \eqref{eopvar2} and \eqref{heestress} we have
\begin{eqnarray}
\Delta S_O=\pi T_{00}\ell^2,
\end{eqnarray}
which is positive. Based on this result we see that the odd entropy increases under the thermal excitations (see section \ref{adsbb}). Of course, this behavior is in qualitative agreement with the earlier QFT results obtained in \cite{Mollabashi:2020ifv}. There the authors considered a two dimensional scalar field theory and found that the odd entropy is an increasing function of the temperature.

We can extend this study to different interesting directions. An interesting question is if either of these behaviors is reflected in the QFT calculations of mixed state correlation measures dual to EWCS. 
Indeed, free field theories provide a particularly well-controlled setup for studying various aspects
of the corresponding measures, \textit{e.g.}, EoP \cite{Bhattacharyya:2018sbw,Camargo:2020yfv} and reflected entropy \cite{Bueno:2020fle,Camargo:2021aiq}. Let us emphasize that to the best of our knowledge, these studies mainly focused on vacuum (Gaussian) states without any excitations or Ising model at the critical point which has conformal symmetry. Notice that the analysis in \cite{Mollabashi:2020ifv} includes the finite temperature corrections to odd entropy in free scalar theories in which mixed states of interest, \textit{e.g.}, thermal states, are again Gaussian. To better clarify the behavior of these measures in generic excited states, it would be interesting to extend these studies to more general setups, \textit{e.g.}, \cite{Alba:2009th,Palmai:2014jqa,Murciano:2018cfp}. This analysis can help us to examine our results and also choose a unique holographic counterpart for EWCS among different proposals. 

In this paper we restricted our discussion to the symmetric configuration for the boundary entangling regions which significantly simplifies the calculation of the EWCS. It is interesting to consider more general configurations where the widths of the strips are different, using the techniques of \cite{Liu:2019qje}. Another topic to explore would include extending our results to field theories dual to higher curvature or massive gravity theories as in \cite{Li:2021rff,Liu:2021rks}. Finally it would be interesting to study EWCS in nonrelativistic holographic setups with nontrivial excitations. We plan to explore some of these directions in the near future.


\subsection*{Acknowledgements}
We would like to thank Mohammad Hasan Vahidinia for ongoing collaboration on related ideas. We would also like to thank Ali Mollabashi for clarifications about aspects of ref. \cite{Mollabashi:2020ifv}. KBV thank the School of Particles and Accelerators, Institute for Research in Fundamental Sciences (IPM) for hospitality where part of this work was carried out.

%
%
%


\end{document}